\documentclass[fleqn,usenatbib]{mnras}

\usepackage{newtxtext,newtxmath}

\usepackage[T1]{fontenc}
\usepackage{ae,aecompl,times}

\usepackage{graphicx}	
\usepackage{amsmath}	
\usepackage{rotating}
\usepackage{orcidlink}

\title[Prompt cusps evolution]{Buried but not destroyed: the evolution from prompt cusps to NFW haloes}

\author[Wang et al.]{Yuchan Wang$^{1} \orcidlink{0009-0003-8612-9971}$,
Sownak Bose$^{1}$ \orcidlink{0000-0002-0974-5266} ,
Carlos Frenk$^{1}$ \orcidlink{0000-0002-2338-716X} \& 
Adrian Jenkins$^{1}$ \orcidlink{0000-0003-4389-2232} 
\\
$^{1}$Institute for Computational Cosmology, Department of Physics, Durham University, South Road, Durham DH1 3LE, United Kingdom}

\date{Accepted XXX. Received YYY; in original form ZZZ}

\pubyear{2025}

\begin{document}
\label{firstpage}
\pagerange{\pageref{firstpage}--\pageref{lastpage}}
\maketitle

\begin{abstract}
The internal structure of dark matter haloes encodes their assembly history and offers critical insight into the nature of dark matter and structure formation. 
Analytical studies and high-resolution simulations have recently predicted the formation of `prompt cusps' - steep power-law density profiles that emerge rapidly from the monolithic collapse of primordial peaks. Yet, by $z=0$ most haloes are well described by Navarro-Frenk-White (NFW) density profiles, raising the question of how these early cusps are transformed in a cosmological setting.
We address this problem using 64 zoom-in $N$-body simulations of eight haloes, each resimulated with eight different free-streaming wavenumbers to control the abundance of small-scale structure while keeping the large-scale environment fixed.
To mitigate numerical discreteness effects, we classify artificial fragments and genuine subhaloes with a physically motivated procedure based on matching subhaloes to their progenitor peaks.
At the population level, we demonstrate that haloes initially form prompt cusps, and their profiles subsequently transition towards the NFW form. 
Our study reveals three distinct evolutionary pathways by which prompt cusps evolve: major mergers, accretion of artificial fragments, and interactions with large-scale filaments, each having a distinct impact on the inner density profile.
In particular, we show that the original power-law cusp remains visible in the profile of particles associated with the primordial peak even when the total halo profile is already NFW-like.
This work highlights the imprint of early collapse on present-day halo structure and provides new insights into the origin of the universality of the NFW profile. 
\end{abstract}




\section{Introduction}

Despite its pivotal role in the $\Lambda$CDM paradigm, the fundamental nature of dark matter remains one of the most important open questions in cosmology.  
While the large-scale distribution of dark matter can be accurately described with a perturbation approach \citep{2002Bernardeau}, the formation of structure on non-linear scales beyond shell crossing remains theoretically challenging.
Numerous studies - spanning analytical models, $N$-body simulations, and observations - have been dedicated to unravelling the formation and evolution of dark matter haloes over the past several decades \citep{2012Frenk}.

Theoretical attempts to explain halo density profiles range from maximum-entropy arguments \citep{1967Lynden-Bell, 2013Pontzen}, and phase-space-based hydrostatic equilibrium models \citep{2013Pontzen}, to post-collapse Lagrangian perturbation theory \citep{2015Colombi}. One of the most influential frameworks was the secondary-infall model, first formulated by \citet{Gunn1972}, which described dark matter halo formation as the radial infall of mass shells onto a central overdensity in an otherwise uniform universe. In this foundational description, spherical concentric mass shells initially expand with the Hubble flow, then reverse their motion and collapse under gravity toward a central point-mass perturbation.
The dynamics of each shell in the Einstein–de Sitter universe are governed by the Newtonian equation of motion:
\begin{equation}
\label{eq:eom}
    \frac{d^2r_{\rm M}}{dt^2} = -\frac{GM(<r_{\rm M}, t)}{r_{\rm M}^2},
\end{equation}
where $r_{\rm M}$ is the radius of the mass shell, $t$ is the time and $M(<r_{\rm M}, t)$ is the mass enclosed inside the shell $r_{\rm M}$ at time $t$. 
\citet{Gott1975} later estimated the resulting density profile by assuming that mass shells stall at their turnaround radius - where they spend the most time - leading to a power-law density distribution,
\begin{equation}
\label{eq:power-law}
    \rho \propto r^{\lambda},
\end{equation}
where $r$ is the distance to the centre in the halo and $\lambda = -9/4$ is an upper limit in this idealised, non-shell-crossing scenario.

Extending the analysis towards the stabilised state long after the first turnaround, \cite{Fillmore1984} and \cite{1985Bertschinger} solved the equation of motion (Eq.~\ref{eq:eom}) for mass shells by applying a rescaling of the time with respect to \textit{its collapse time}, enabling a self-similar description of all mass shells.
For the self-similar solutions to be reached, the initial perturbation must be scale-free, meaning the enclosed mass is described by a power law with the distance towards its centre.
The numerical integration of the equation of motion suggested that the trajectory of the mass shells stabilises after a few orbits \citep{1985Bertschinger}, yielding a central density profile that remains unchanged over time, with a slope directly linked to the initial conditions.
Unlike the isothermal profile with $\lambda = -2$ that results from the violent relaxation picture in \cite{1967Lynden-Bell}, the self-similar solution preserves the memory of the initial density distribution.

Recently, \citet{White2022} introduced a formulation that describes the early collapse of density peaks into `prompt' cusps. By rescaling the trajectories of mass shells {relative to the time difference between their collapse and that of the central peak, \citet{White2022} derived a self-similar solution that yields a steep inner density profile with a power-law slope of approximately $-12/7$.
This prompt cusp forms rapidly following the collapse of the central region and is expected to persist through later stages of halo evolution, until it becomes embedded within a broader density structure once the self-similar conditions break down.
This idealization - particularly the requirement of an isolated peak with a power-law density profile - limits its direct applicability to realistic cosmological haloes and $N$-body simulations, where departures from spherical symmetry and hierarchical growth complicate the internal dynamics.

In realistic cosmological environments, deviations from internal spherical symmetry and the presence of external tidal fields lead to departures from idealized, self-similar collapse.
\citet{White2022} emphasizes that the deviations from spherical symmetry of the internal structure of the peak itself can lead to a spread in collapse times along the three principal axes. This anisotropy introduces ambiguity into the self-similar rescaling procedure, which assumes a unique and consistent collapse time for the central peak.
Nearby inhomogeneities induce an anisotropic tidal field, which develops a non-radial velocity component in the motion of the infalling mass shells \citep{1969Peebles}. 
However, analytical treatments that incorporate angular momentum while preserving self-similarity - such as those by \citet{1992White, 1994White} - have shown that the $-12/7$ index will remain unaffected. This finding is supported by later work, including \citet{2001Nusser} and \citet{2004Ascasibar}.
Other analytical work suggests that using a classical form of specific angular momentum flattens the power-law density profile, albeit at the cost of violating self-similarity \citep{2023Popolo}.
Analytical models based on isolated peaks fail to capture the full complexity of hierarchical structure formation in a cosmological context.
Subsequent mass accretion from regions beyond the initial peak breaks the self-similar conditions assumed in idealized models.

In cosmological simulations \citep{1996Cole} and zoom-in studies \citep{2008Springel}, haloes are found to consistently evolve toward an equilibrium state: the Navarro–Frenk–White (NFW) profile \citep{1996Navarro, 1997nfw}. The spherically averaged mass density profile of an NFW halo is described by two parameters, 
\begin{equation}
    \rho(r) = \frac{\rho_s r^3_s}{r(r + r_s )^2},
\end{equation}
where $\rho_s$ and $r_s$ are the characteristic density and scale radius, respectively. The profile asymptotes to $\rho \propto r^{-1}$ at small radii and $\rho \propto r^{-3}$ at large radii. 
A remarkable feature of the NFW profile is its universality: it provides a good description of haloes across a wide range of formation histories and in different cosmological models \citep{1999Huss, 2009Wang}. This universality has been confirmed over more than 20 orders of magnitude in halo mass at redshift zero \citep{Wang2020}, down to the free-streaming cut-off scale expected for a 100 GeV WIMP ($\sim 10^{-6} M_\odot$), although simulations altering the amplitude and shape of the primordial power spectrum indicate that halo profiles may deviate from NFW under non-standard initial conditions \citep{2020Brown}.
Although the Einasto density profile provides a slightly better fit with an additional shape parameter \citep{2004Navarro}, resulting in a departure from the strict universality \citep{2010Navarro}, the regularity of halo structure remains of theoretical interest, particularly because it extends beyond density alone \citep{2005Austin}: related quantities such as phase-space density \citep{2001Taylor} and angular momentum distributions \citep{2001Bullock} also show universal behaviour, despite the diversity seen in the mass accretion history.

Low-mass haloes near the free-streaming cut-off offer a valuable opportunity to test the role of hierarchical clustering in the formation of the NFW profile.
Their sparse merger histories increase the likelihood that their initial collapsed structure remains intact, and also make them well-suited for isolating the impact of individual mergers - especially for simulations with limited time resolution.

The formation of prompt cusps has been observed in various high-resolution simulations as an early stage in halo evolution. For example, \citet{2005Diemand} found that the density profiles of Earth-mass ($10^{-6}\, M_\odot$) haloes at high redshift devoid of substructures follow a single power-law with slopes ranging from $-1.5$ to $-2$. However, their simulation was halted at $z=26$ due to the distortion caused by the complex tidal field generated by the surrounding low-resolution region, illustrating the challenges in resolving such small scales across a large dynamical range.
Cosmological simulations with increased resolution have statistically showed that small haloes at high redshift typically exhibit inner slopes steeper than NFW, with values between $-1.4$ and $-1.5$ \citep{2010Ishiyama, 2014Ishiyama}. 
Other studies have adopted manually modified power spectra to control structure formation at specific scales \citep{2018Delosb, 2019Delos}. 

Warm Dark Matter (WDM) simulations have been used to model structure formation with truncated power spectra, effectively acting as "scaled-up" analogues of Earth-mass haloes in cold dark matter (CDM) cosmologies \citep{2008Colin, 2014Ishiyama}. Yet these approaches face complications from artificial processes, where artificial fragmentation of filamentary structures arises from discreteness noise in the initial particle distribution \citep{Wang2007}. These artificial objects not only contaminate the halo population \citep{Wang2007} but can also influence the evolution of genuine structures \citep{Ondaro2024}, as their interactions - such as merging - can mimic those of real small-scale haloes, underscoring the critical need to identify and mitigate numerical artefacts in simulations studying prompt cusp formation.

One approach to suppress artificial fragmentation is to improve mass and force resolution. This can be achieved via targeted zoom-in simulations of individual haloes \citep{2013Anderhalden, 2018Delos}. In WDM simulations, \citet{2013Anderhalden} found that haloes exhibited inner density slopes of $-1.3$ to $-1.4$, in contrast to the NFW-like profiles that formed in the CDM control runs. 

Finally, idealised simulations can be used to test individual mechanisms in controlled settings. For example, \citet{2018Ogiya} studied mergers of cored proto-haloes and found that the resulting profile consistently evolved toward a power-law slope of $-1.5$, independent of large-scale non-spherical perturbations. the idealised setup omits environmental factors such as tidal torques from nearby filaments. 
 
 In addition to studies on the slope of the inner density profile, \citet{2019Delos} developed a model linking the properties of individual haloes to those of their progenitor peaks. This model predicts the amplitude and cosmological distribution of halo density profiles to within 20\% accuracy using only parameters derived from the initial conditions - the peak height, characteristic size, and local tidal tensor - which are largely independent of the underlying cosmology \citep{2023Delos, Ondaro2024}. Notably, these predictions are also consistent with the analytical amplitude estimates from \citet{White2022}, with deviations within approximately 40\%.

The evolution of prompt cusps and their relationship to the NFW-like profiles observed in simulations has been explored with recent high-resolution $N$-body simulations. 

The prompt cusp provides an useful baseline for testing how various mechanisms - major mergers, smooth accretion, and external disturbances - affect inner density structure. As with the broader debate surrounding the origin of the NFW profile, there is no consensus regarding the dominant mechanisms that disrupt the prompt cusps. \citet{2014Ishiyama} suggested, based on stacked profiles at high redshift, that frequent major mergers near the free-streaming scale could lead to a shallowing of the central cusp via enhanced central densities. This trend was also seen in cosmological simulations by \citet{2019Delos}.
\citet{2016Ogiya} tested this hypothesis using idealized simulations with controlled mergers. They found that repeated major mergers can significantly flatten the inner slope of a prompt cusp, with moderate dependence on progenitor orbital parameters. In contrast, mergers between NFW haloes have a more uniform density increase across radii, thus do not substantially alter the slope. This asymmetry suggests a possible evolutionary pathway: while prompt cusps may flatten toward the $r^{-1}$ slope through violent relaxation, established NFW haloes appear more resilient to further mergers, leading to an erasure of the imprint of the primordial peak, resulting in an universal profile.
In targeted zoom-in simulations of haloes within cosmological environments, \citet{2017Angulo} observed a transition in the inner slope from $-1.5$ to $-1$ with frequent mergers, while smooth accretion was found to preserve the central density. However, they also noted that the final profiles deviated from the NFW form, likely because the simulation ended before the haloes reached full equilibrium.

In a recent study, \citet{2023Delos} explored the relationship between prompt cusps and their progenitor peaks by capturing the cusps at formation and tracing their evolution until they transitioned into NFW profiles.
In contrast to earlier findings, mergers were found to have only moderate influence on cusp evolution, whereas accretion played a more dominant role. Accretion was shown to add mass at intermediate radii while leaving the central cusp intact, effectively shallowing the profile.

While mergers and accretion represent different aspects of mass assembly, together they contribute to both redistributing material from the original prompt cusp and accumulation new mass from the surroundings.
In this work, we aim to disentangle these two effects by identifying the matter that originated from the progenitor peak that collapsed early in accordance with self-similar predictions, and separating it from material accreted from the surroundings.
By tracing the spatial distributions of these two components, we gain deeper insight into the mechanisms behind the apparent "destruction" of prompt cusps. In particular, this allows us to distinguish between genuine structural disruption and the mere embedding of the original cusp within later-accreted material, both of which can alter the measured density profile slope.

Furthermore, studying the formation and evolution of prompt cusps is of importance from both theoretical and observational perspectives.
One widely studied dark matter candidate is the weakly interacting massive particle (WIMP; \citealt{1996Jungman}). As a byproduct of the inverse process of pair production in the early universe, WIMP annihilation can produce high-energy $\gamma$-ray photons, with luminosity directly tied to the distribution of dark matter substructure (\citealt{1998Bergstr, 2008Springel}). These annihilation signals may be observed in several forms: as an isotropic $\gamma$-ray background (\citealt{2015Ackermann}), as excess emission from the Galactic centre (\citealt{2011Hooper, 2017Ackermann}), or as individual point sources (\citealt{2012Ackermann}).
An ideal dark matter model must account for both the detections and non-detections of annihilation signals across the various observational channels \citep{2012Bringmann}. Such a model must describe both the internal density structure of dark matter (sub)haloes and their abundance, accounting for hierarchical formation and baryonic feedbacks.
Given their steep inner density slope ($\rho \propto r^{-12/7}$) and potential survival in cosmological environments (\citealt{Delos2023, 2023Delos}), prompt cusps may contribute significantly to WIMP annihilation signals.
\footnote{The divergence of density is constrained by the Liouville’s theorem \citep{1979Tremaine}, resulting in a core at the centre with maximum phase-space density set by the primordial velocity dispersion \citep{Delos2023}. }

Signal predictions from prompt cusps - whether as isolated point sources (\citealt{2024Delos}), diffuse emissions from the Galactic centre (\citealt{2018Delos, 2023Delos}) and distant clusters (\citealt{2025Crnogor}), or the isotropic background (\citealt{2024Delosb, 2024Ganjoo}) - consistently show that, despite their low total mass fraction, prompt cusps dominate the annihilation luminosity budget. An exception occurs in the Galactic centre, where tidal disruption by stellar encounters suppresses their contribution (\citealt{2022Stucker}).
Moreover, high-resolution simulations by \citet{2023DelosD} show that in warm dark matter (WDM) models, prompt cusps can reach masses of order $10^7,M_\odot$, potentially altering the observed kinematics of Local Group dwarf galaxies.

In this work, we investigate how prompt cusps that arise from the collapse of primordial density peaks evolve into NFW-like haloes, with particular emphasis on how different mass-assembly channels and environmental perturbations modify the inner density profile. We quantify the response of the density distribution to smooth accretion, minor and major mergers, and to numerical artefacts in the form of artificial fragments. Section~\ref{sec:method} describes the simulation setup and the selection of our zoom-in targets (Section~\ref{subsec:halo_selection}), including the construction of truncated-power spectra that vary the free-streaming scale while keeping the large-scale environment fixed. In Section~\ref{subsec:spurious} we develop and calibrate a physically motivated diagnostic to identify artificial fragments, and provide broadly applicable mass-shape selection criteria that remove them with controllable completeness and purity. Section~\ref{subsec:self-similar} outlines our peak-based framework for separating particles originating from the self-similar collapse of the primordial peak from material accreted later, and for constructing an improved self-similar description that explicitly incorporates the Lagrangian peak shape. In Section~\ref{sec:example_merger} - ~\ref{sec:example_spurious} we step through three representative evolutionary pathways in detail - a case dominated by a major merger, a case of interaction with a filament with relatively quiescent accretion, and a case where an early merger with an artificial fragment strongly perturbs the central structure - to illustrate concretely how the mechanisms identified in the statistical analysis operate in individual haloes. The global evolution of the halo and peak-only density profiles, and their comparison with NFW and various power-law prompt–cusp models, is then presented in Section~\ref{subsec:fitting}.

\section{Halo simulations}
\label{sec:method}
\begin{figure}
    \centering
    \includegraphics[width=\columnwidth]{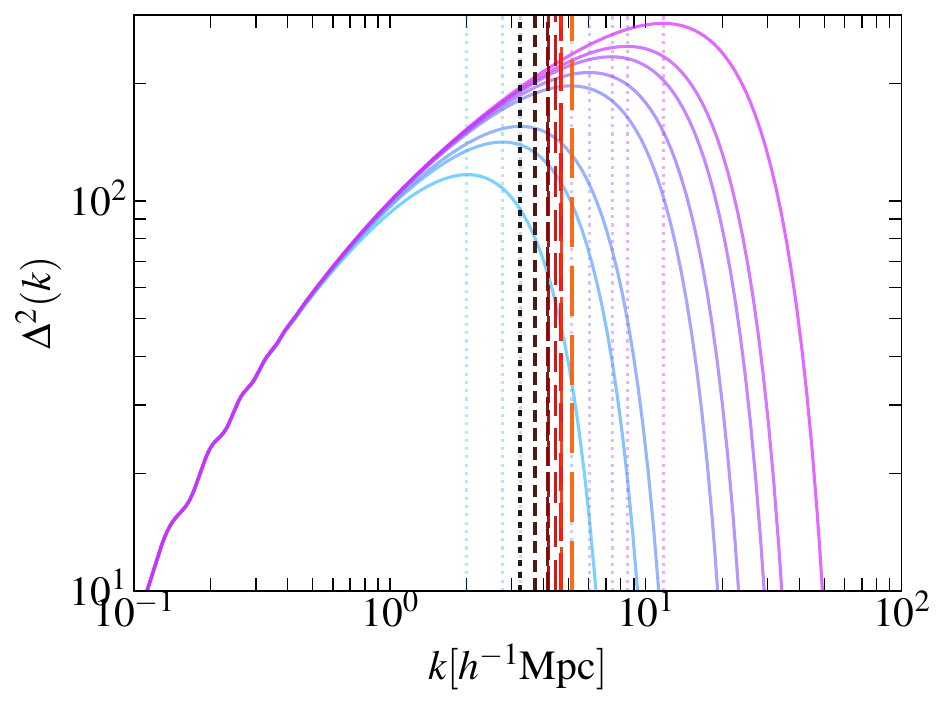}
    \caption{
    Dimensionless power spectrum $\Delta^2(k)$ for halo zoom-in simulations with varying small‐scale suppression scales $k_{\rm fs}$ (solid curves). Dotted lines mark the peak wavenumber, $k_{\rm max}$, for each power spectrum, while dashed lines indicate the characteristic scale, $k_{\rm c}$, of eight haloes, ranked from least to most massive. Both axes are shown in logarithmic units.}
    \label{fig:IC_PS}
\end{figure}

At scales near the free-streaming cut-off, the initial matter distribution becomes sufficiently smooth to isolate the collapse of a single peak. Since structure formation is suppressed below this scale, these conditions allow us to capture the moment of halo formation - just after collapse, but before significant accretion or merging occurs.

Our choice of initial conditions is motivated by the need to isolate the collapse and early evolution of individual primordial peaks while controlling the degree of subsequent hierarchical growth. By systematically varying the small-scale cut-off in the power spectrum, we mimic different free-streaming scales and merger histories without altering the large-scale environment of each target halo. This approach allows us to capture prompt cusp formation under conditions close to the idealised self-similar collapse models, while also enabling controlled tests of how environmental perturbations, mergers, and smooth accretion influence the transition to the NFW profile.

We construct a suite of 64 warm dark matter simulations by re-simulating 8 haloes of varying mass, $M$, each under 8 different free-streaming cut-off scales, $k_{\rm fs}$, in the initial power spectrum. These cut-offs are chosen to span a range that brackets the characteristic scale of each halo, approximated by the extent of its Lagrangian region:
\begin{equation}
\label{eq:halo_kc}
    \bar{\rho} 4 \pi r_{\rm ini}^3 = M,
\end{equation}
where $\bar{\rho}$ is the mean background density, and $r_{\rm ini} = \pi / k_{\rm fs}$ defines the characteristic radius associated with the peak. The values of $k_{\rm fs}$ explored in this work are $[7,\,10,\,12,\,20,\,24,\,30,\,35,\,50]\,h\,\mathrm{Mpc}^{-1}$.

All haloes are selected from the Level 2 Voids-in-Voids-in-Voids (VVV) simulation suite \citep{Wang2020}, which adopts the following cosmological parameters:
\begin{equation}
    \{\Omega_m,\Omega_\Lambda,h,\sigma_8\} = \{0.307,0.693,0.677,0.829\},
\end{equation}
where $\Omega_m$ and $\Omega_\Lambda$ denote the matter and dark energy density fractions, $h$ is the dimensionless Hubble parameter, and $\sigma_8$ is the rms amplitude of matter fluctuations on $8\,h^{-1}\mathrm{Mpc}$ scales.

The same linear power spectrum as in the parent simulation is used, modified by a damping function to account for small-scale suppression in WDM:
\begin{equation}
    D(k) = (1 - \frac{2}{3}\left(\frac{k}{k_{\rm fs}}\right)^2)\exp({-\left(\frac{k}{k_{\rm fs}}\right)^2}),
\end{equation}
following \citet{Green2004}.
The corresponding dimensionless power spectrum $\Delta^2(k)\propto D^2(k) $ for different $k_{\rm fs}$ are shown in Figure~\ref{fig:IC_PS}, with larger $k_{\rm fs}$ values exhibiting less suppression on small scales. For example, $k_{\rm fs} = 7\,h\,\mathrm{Mpc}^{-1}$ heavily suppresses the formation of the most massive halo (leftmost dashed line), whereas $k_{\rm fs} = 50\,h\,\mathrm{Mpc}^{-1}$ represents a weak suppression scenario, effectively mimicking cold dark matter (CDM) behaviour at these scales.

By varying $k_{\rm fs}$, we systematically alter each halo’s merger history and properties of their primordial peak while preserving its large-scale environment. This dual effect allows us to study two key stages of prompt cusp evolution: (1) how changes in the initial peak shape affect the formation of the prompt cusp, enabling direct comparison to theoretical predictions such as those in \citet{White2022}; (2) how subsequent interactions, such as mergers, affect its stability or destruction.

\subsection{Halo selection}
\label{subsec:halo_selection}
The original Voids-in-Voids-in-Voids (VVV) simulation suite employs a recursive zoom-in technique, resimulating progressively smaller nested regions from low resolution (Level 0) to high resolution (Level 8). Each refinement level achieves higher particle mass resolution within its high-resolution zone. However, structures located near the boundaries of these zones can suffer from numerical artefacts, particularly tidal heating from coarse background particles. Consequently, some haloes identified in the parent simulation by Friends-of-Friends (FOF; \citealt{Davis1985}) or \texttt{SUBFIND} \citep{Springel2001} are disrupted or entirely lost when resimulated.

To minimise such contamination, we select haloes based on the Lagrangian properties of their progenitor regions in the parent simulation. We define four diagnostic parameters:
\begin{itemize}
    \item Effective Lagrangian radius ($r_{\rm L}$) – the mean radial extent of the Lagrangian volume in the initial conditions, measuring the average spread of particles in Lagrangian space. In the absence of particle mixing, the Lagrangian volume should correspond one-to-one with the final halo mass.
    \item Minor-to-major axis ratio ($[c/a]_{\rm L}$) – quantifies deviations from spherical symmetry of the Lagrangian region. We follow the elliptical fitting procedure of \citet{Schneider2012}, computing the reduced inertia tensor:
    \begin{equation}
    I_{ij} = \sum_{n=1}^{N} \frac{x_{n,i} x_{n,j}}{R_n^2} ,
    \end{equation}
    where $x_{n,i}$ are the Lagrangian coordinates relative to the region’s centre and $R_n$ is the elliptical radius to the centre of the primordial peak:
    \begin{equation}
    R_n^2 = \sum_{i=1}^{3} \frac{e_{n,i}^2}{a_i^2} ,
    \end{equation}
    with $e_{n,i}$ the coordinates along the principal axes and $a_i$ the axis lengths. In our simplified implementation, $a_i$ is initialised to half the maximum extent of the region, $e_{n,i} = x_{n,i}$, and the iteration stops when $[c/a]_{\rm L}$ changes by less than 5\%. Weighting by $R_n^{-2}$ suppresses the influence of outlying particles.
    \item Fraction of external high-resolution particles ($N_{\rm L,o}$) – the fraction of high-resolution particles lying inside the Lagrangian region but not belonging to the halo at $z=0$. This metric captures contamination from particles that will become high-mass intruders in the resimulation.
    \item Spearman rank correlation ($C_{\rm sp}$) – measures the monotonic correspondence between each particle’s initial radial distance from the Lagrangian centre and its final distance from the halo centre at $z=0$. Under the self-similar assumption, the movements of all mass shells follow similar trajectories when rescaled by appropriate parameters \citep{White2022}. \citet{Delos2023} found that, in this regime, the distribution of particles inside the halo mirrors their distribution in the Lagrangian region, whereas mergers disrupt this correspondence. Therefore, $C_{\rm sp}$ provides a quantitative measure of the orderliness of particle distributions, allowing us to monitor a halo’s merger history. This serves as a complementary diagnostic to the mass accretion history, which encodes the combined effects of smooth and clumpy (merger-driven) accretion. A high $C_{\rm sp}$ indicates an ordered, self-similar collapse with minimal disruption, while low values signal a disturbed dynamical history. The Lagrangian centre is defined as the location of the main density peak in the smoothed linear density field, not the geometric centroid of the particle distribution.
\end{itemize}

We construct an initial sample of haloes from VVV Level 2 containing more than $10^3$ particles. From this, we select candidates by applying percentile thresholds to the three Lagrangian quality metrics ($r_{\rm L}$, $[c/a]_{\rm L}$, $N_{\rm L,o}$), requiring all selected haloes to fall within the lowest $m\%$ of the parent population for each criterion. This ensures minimal boundary interference and good convergence in resimulations. To span a range of merger histories, we then split the final sample by selecting four haloes from the lower $m\%$ of the $C_{\rm sp}$ distribution and four from the upper $m\%$. The value of $m$ is tuned (between $15\%$ and $20\%$) to yield exactly eight haloes satisfying all criteria. The resulting sample spans halo masses from $2.1 \times 10^9 \mathrm{M}_\odot$ to $1.5 \times 10^{10} \mathrm{M}_\odot$ and is summarised in Table~\ref{tab:Halo_params}.

\subsection{Simulation setup}
For each selected halo, we generate initial conditions (ICs) using the procedure in \citet{Wang2020}. 
We first construct a rectangular mask in Lagrangian space to closely encompass  the irregular halo Lagrangian region in the parent simulation. Cells inside the Lagrangian volume are labelled as `high' resolution, adjacent buffer zones as `fine', and the remainder as `coarse'. Each region is then populated with particles at different grid resolutions: $20^3$ for high, $15^3$ for fine, and $10^3$ for coarse cells. The mass distribution is centred within each grid cell and initialized to the cosmic mean density. Increasing or decreasing the number of particles inside `fine' and `coarse' cells does not result in visible difference in the simulation results.
To reproduce the large-scale tidal field, the zoom-in region is embedded within 99 concentric cubic shells of increasing particle mass, each containing $(n+2)^3 - n^3$ particles (where $n$ is the grid size of the shell), centred on the mask box. In this work, $99$ shells of increasing sizes are used to fill a periodic simulation box of $542.16 \ h^{-1}{\rm Mpc}$.

Initial particle displacements and velocities were generated at redshift \(z_{\rm ini}=255\) with \verb|IC_Gen| \citep{Jenkins2013} using second-order Lagrangian perturbation theory (2LPT). For each target halo, we produced eight otherwise identical sets of initial conditions that differ only in their input linear matter power spectrum, thereby imposing distinct small-scale cut-offs (see Section~\ref{sec:method}).

The 64 initial conditions are then evolved forward with \verb|Gadget-4| \citep{2021Springel}. We enable individual gravitational softening lengths that scale with particle mass, following \citet{Power2003}:
\begin{equation}
    \epsilon = \frac{4r_{200}}{\sqrt{M_{200}/M_{\rm p}}}
\end{equation}
where $M_{\rm p}$ is the mass of the constituent particles of the target halo. 

\begin{table*}
\centering
\begin{tabular*}{\textwidth}{@{\extracolsep{\fill}}c|c|c|c|c|c|c|c|c|c|c}
\hline\hline
Halo ID & $\mathrm{M}_{\rm 200}[\mathrm{M}_{\odot}]$ & $r_{\rm 200}[h^{-1}{\rm Mpc}]$ & $\epsilon[h^{-1}{\rm Mpc}]$  & $m_{\rm p}[\mathrm{M}_{\odot}]$ & $N_{\rm p}$ & $k_{\rm ini}$ [$h\ {\rm Mpc}^{-1}$]  \\
\hline
108 & $2.04\times 10^{10}$ & $4.45\times 10^{-2}$ & $6.86\times 10^{-5}$ & $3.04\times 10^{3}$ & $6.71\times 10^{6}$ & 11.76\\ 
206 & $9.71\times 10^{9}$ & $3.47\times 10^{-2}$ & $5.93\times 10^{-5}$ & $1.77\times 10^{3}$ & $5.48\times 10^{6}$ & 15.06\\
305 & $6.81\times 10^{9}$ & $3.01\times 10^{-2}$ & $5.58\times 10^{-5}$ & $1.39\times 10^{3}$ & $4.88\times 10^{6}$ & 16.96\\
443 & $5.04\times 10^{9}$ & $2.79\times 10^{-2}$ & $6.61\times 10^{-5}$ & $1.77\times 10^{3}$ & $2.84\times 10^{6}$ & 18.74\\
\hline
150 & $1.38\times 10^{10}$ & $3.90\times 10^{-2}$ &  $8.51\times 10^{-5}$ & $4.11\times 10^{3}$ & $3.36\times 10^{6}$ & 13.40\\
225 & $9.63\times 10^{9}$ & $3.46\times 10^{-2}$ & $4.81\times 10^{-5}$ & $1.16\times 10^{3}$ & $8.27\times 10^{6}$ & 15.11\\
260 & $8.01\times 10^{9}$ & $3.25\times 10^{-2}$ & $5.77\times 10^{-5}$ & $1.57\times 10^{3}$ & $5.10\times 10^{6}$ & 16.06\\
337 & $6.70\times 10^9$ & $3.07\times 10^{-2}$ & $3.79\times 10^{-5}$ & $6.39\times 10^{2}$ & $1.05\times 10^{7}$ & 17.05\\

\hline\hline

\end{tabular*}
\caption{The numerical parameters and $z=0$ halo properties for each resimulated halo in the parent simulation. {\it Column 1}: the halo index in the parent simulation. {\it Column 2}: the virial mass of the halo in the parent simulation. {\it Column 3}: the virial radius of the halo in the parent simulation. {\it Column 4}: the softening length used in the halo resimulations. {\it Column 5}: the mass resolution of the particles in the resimulated halo. {\it Column 6}: the number of particles inside the halo in the halo resimulations. {\it Column 7}: the characteristic wavelength corresponding to the halo size $r_{\rm ini}$ in the initial condition (Eq.~\ref{eq:halo_kc}).
The 4 haloes listed in the top 4 rows are the most ordered haloes selected by constraining both the Lagrangian parameters and $C_{\rm sp}$ to be in the lowest $15\%$ of the entire sample, while the 4 haloes at the bottom rows are chosen with the reversed $C_{\rm sp}$ constraint (Section~\ref{subsec:halo_selection}), which means that they are among the most disordered haloes.}
\label{tab:Halo_params}
\end{table*}

\section{Method}
\subsection{Artificial fragmentation}
\label{subsec:spurious}
Artificial fragmentation, or spurious haloes, are not physical structures seeded by primordial perturbations, but artefacts that arise from discreteness noise in $N$-body simulations. These structures typically appear along filaments as clumpy overdensities. Their typical mass has been found to scale as $k_{\rm max}^{-2}$, where $k_{\rm max}$ is the wavenumber at which the power spectrum $\Delta^2(k)$ reaches its maximum \citep{Wang2007}. Since our resimulated haloes are constructed near the free-streaming cut-off of the power spectrum, the presence of artificial fragmentation is expected.

\citet{Ondaro2024} studied the impact of artificial fragmentation on prompt cusps by comparing standard $N$-body simulations with phase-space-based approaches \citep{2013Hahn, 2016Hahn}, where the mass is smoothly distributed over phase-space elements, suppressing discreteness effects. They found that artificial structures tend to shallow the inner density slope of prompt cusps. Motivated by this, we treat artificial fragmentation as one of the mechanisms that may disturb or destroy prompt cusps, along with other evolutionary pathways (see Figure~\ref{fig:Abstract}).

\subsubsection{Identifying artificial fragmentation}
\label{subsubsec:artificial_identification}
To study these effects systematically, we identify artificial structures both prior to and after the formation of prompt cusps.
A widely used empirical method is that of \citet{Wang2007} and \citet{Lovell2014}, which identifies artificial fragments  based on the combination of their mass, the shape of the Lagrangian patch from which they originate, and the mass resolution of the simulation in question. The mass threshold delineating real and spurious haloes is given by:
\begin{equation}
\label{eq:M_lim}
    M_{\rm lim} = \kappa10.1 \Bar{\rho} d k^{-2}_{\rm max},
\end{equation}
where $\bar{\rho}$ is the mean density of the universe, $d$ is the inter-particle spacing (i.e. grid resolution of the high-resolution region), and $\kappa$ is a calibration constant. This threshold is calibrated to ensure that subhaloes appearing in both low- and high-resolution simulations are preferentially classified as genuine.

In contrast to this empirical strategy, we develop a physically motivated approach based on whether a given (sub)halo originates from a genuine primordial peak in the initial conditions. Real haloes arise from gravitational instability seeded by peaks in the linear density field, while artificial fragments do not.
The primordial dark matter distribution in our simulations is sufficiently smooth to allow precise and complete identification of primordial peak locations using a straightforward method.
We start from random positions in the Lagrangian region of each main halo and perform steepest-ascent walks along the grid to locate density peaks. Each endpoint is marked as a local maximum, corresponding to a primordial peak. This procedure is repeated for multiple initial positions until all distinct peaks are identified.
No additional smoothing is required, since the truncated power spectrum naturally suppresses small-scale fluctuations. 

For each identified peak, we compute its peak height, $\delta_{\rm p}$, defined as the local maximum of the linear overdensity field. 
From the peak height, we derive the corresponding peak scale, 
$r_{\rm p} \equiv |\delta / \nabla^2 \delta|$, 
which characterizes the typical spatial extent of the peak. 
This scale determines the set of particles encompassed within the neighbourhood of the peak. 
We note that the precise definition of the peak boundary is not critical for the subsequent analysis, as the outer regions contribute only weakly to the overdensity compared to the peak centre and are effectively down-weighted in the following calculations.

We then trace each subhalo in the simulation back to the snapshot when it reaches its maximum pre-infall mass, $M_{\rm sub}$. Bound particles within the proto-subhalo are identified and traced back to their Lagrangian positions, which define the subhalo's Lagrangian region, and the set of relevant particles.

To match subhaloes to peaks, we adopt a modified version of the structural overlap statistic from \citet{Lovell2014}, defined as:
\begin{equation}
    S = \frac{U^2_{\rm ph}}{U_{\rm pp}U_{\rm hh}}, 
\end{equation}
where $S$ is the score characterizing the degree of the overlap between the two regions, ${\rm p}$ and ${\rm h}$, and the cross- and self-correlation terms are:
\begin{equation}
    U_{\rm xy} = \sum_{V_{\rm x}\cup V_{\rm y}}\frac{\delta_{\rm x}\delta{\rm y}}{\sqrt{d_{\rm xy}^2 + s^2}},
\end{equation}
where the subscripts ${\rm x}$ and ${\rm y}$ refer to either the primordial peak (${ \rm p }$) or the subhalo (${ \rm h }$), $d_{\rm xy}$ is the separation between particle pairs, and $s^2$ is a small constant introduced to ensure numerical stability. 
The structural overlap statistic, $S$, provides a quantitative measure of how strongly the Lagrangian proto-region of a subhalo overlaps with that of a density peak. 
In practice, $S$ compares the cross-correlation between the density-weighted particle distributions of the two regions to their respective self-correlations, such that it is normalized to unity for perfect correspondence. 
Values of $S$ therefore range from $0$ to $1$, where $S=1$ indicates complete spatial overlap between the subhalo and peak regions, while $S=0$ corresponds to no spatial correlation. 
A spurious subhalo will have negligible overlap with any nearby peak, resulting in $S \simeq 0$. 
Due to pre-processing effects that can alter a halo before its infall, as well as the complex, non-spherical collapse of the primordial peak, the overlap score $S$ between a subhalo and its progenitor is typically less than 1.

In the practical matching procedure, we require that each peak with a height exceeding the current spherical collapse threshold be assigned to the halo that yields the highest overlap score, while ensuring that no peak or halo is matched more than once.
We establish a one-to-one correspondence for a subset of subhaloes whose Lagrangian regions overlap the most significantly with identified peaks. These subhaloes are classified as physical, while those lacking any matched peak are considered to have formed unphysically and are thus identified as artificial fragments. This method complements conventional criteria based on mass and Lagrangian shape (e.g. axis ratio), providing an independent calibration that avoids the need for additional comparison simulations at different resolutions \citep{Lovell2014}.

For each simulation, we compute a binary field identifying whether a subhalo is genuine or artificial, and evaluate a set of standard detection metrics across a uniform two-dimensional grid of thresholds applied to the maximum progenitor mass normalized by the artificial mass limit, $\log_{10}(M_{\rm sub}/M_{\rm lim})$, and the Lagrangian elliptical axis ratio $c/a$.
At each point of the $100\times100$ grid, the following quantities are measured:
\begin{equation}
\mathrm{TPR} = \frac{N_{\mathrm{TP}}}{N_{\mathrm{TP}} + N_{\mathrm{FN}}}, 
\qquad
\mathrm{FPR} = \frac{N_{\mathrm{FP}}}{N_{\mathrm{FP}} + N_{\mathrm{TN}}},
\label{eq:tpr_fpr}
\end{equation}
where $N_{\mathrm{TP}}$, $N_{\mathrm{FP}}$, $N_{\mathrm{TN}}$, and $N_{\mathrm{FN}}$ denote the numbers of true positives, false positives, true negatives, and false negatives, respectively. 
The true positive rate (TPR) quantifies the completeness of genuine detections, whereas the false positive rate (FPR) measures the contamination from artificial fragments mistakenly classified as genuine. 

A single scalar measure of the overall discrimination power is provided by Youden's $J$ statistic: \citep{1950Youden}, defined as
\begin{equation}
J = \mathrm{TPR} - \mathrm{FPR},
\label{eq:youden}
\end{equation}
which ranges from $J = 0$ (no discriminative power) to $J = 1$ (perfect classification). 
This metric balances completeness and purity, and is independent of the relative abundance of real and spurious objects.

The optimal joint threshold is determined by locating the global maximum of $J$ across the two-dimensional threshold grid. 
The resulting thresholds correspond to the parameter combination that simultaneously maximizes the detection rate of genuine subhaloes and minimizes contamination by artificial fragments.

\subsubsection{The distribution of artificial fragments}
\begin{figure*}
    \hspace*{-3cm}
    \centering
    \includegraphics[width = 1.3\textwidth]{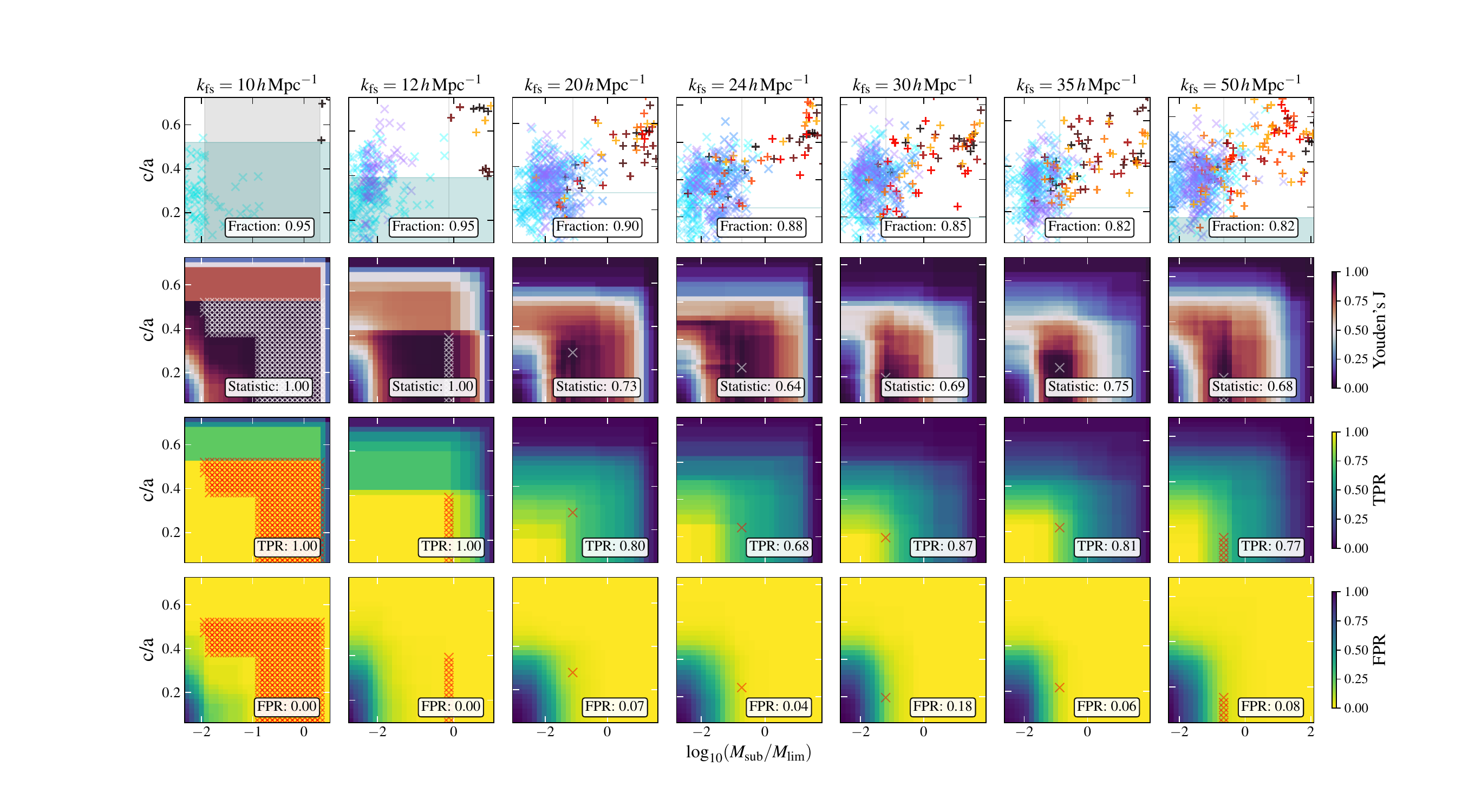}
    \caption{Detection performance of artificial fragments  that appear as subhaloes across different free-streaming scales, with each column corresponding to a set of simulations with a distinct free-streaming wavenumber $k_{\mathrm{fs}}$, as indicated in the title. 
    \textit{Top row}: scatter plots of the maximum progenitor mass of the subhaloes, $M_{\rm sub}$, normalized by the artificial mass threshold $M_{\rm lim}$ \citep{Wang2007}, versus the Lagrangian axis ratio $c/a$ for subhaloes matched (\textit{red, +}) or unmatched (\textit{blue, $\times$}) to density peaks; marker colours indicate the host halo rank. The fraction of spurious subhaloes is annotated in the bottom-right corner of each panel. 
    \textit{Second to fourth rows}: density maps showing Youden's $J$ statistic, the true positive rate (TPR), and the false positive rate (FPR) as functions of joint threshold cuts on $\log_{10}(M_{\mathrm{sub}}/M_{\mathrm{lim}})$ and $c/a$. 
    Grey dashed and dotted lines in the top row, white markers in the second row and red markers in the two bottom rows mark the optimal thresholds for Youden's $J$. And as the optimal threshold spans a finite range in the parameter space, it is indicated in the top panels as shaded regions. 
    Colour bars to the right of each row provide the corresponding metric scales.
    }
    \label{fig:spurious_result}
\end{figure*}
With the identification of artificial fragments described in Section~\ref{subsubsec:artificial_identification}, 
we next examine the distribution and statistical properties of artificial fragments across the suite of simulations with different free-streaming wavenumber $k_{\mathrm{fs}}$. 

Each column in Figure~\ref{fig:spurious_result} corresponds to one simulation set with a distinct $k_{\mathrm{fs}}$, ranging from $10$ to $50~h\,\mathrm{Mpc^{-1}}$. The top row of Figure~\ref{fig:spurious_result} shows the distribution of subhaloes in the parameter space defined by the normalized progenitor mass, 
$M_{\mathrm{sub}}/M_{\mathrm{lim}}$, and the Lagrangian axis ratio, $c/a$. 
Blue points represent subhaloes not associated with any primordial density peak (artificial), whereas red points correspond to genuine subhaloes with an identified progenitor; the fraction of artificial fragments is annotated in each panel.
Marker colours indicate the rank of the subhaloes inside the host halo, with a black cross representing the main subhalo.

In the top panels, the distribution of genuine subhaloes is concentrated in the upper-right region of the parameter space, corresponding to larger $c/a$ values and higher masses. 
In contrast, artificial fragments cluster toward the lower-left region, spanning almost the entire range along both axes. 
This distribution demonstrates that it is impractical to separate artificial and genuine subhaloes using a single parameter alone. 
No clear correlation is observed between $c/a$ and mass for the artificial fragments, as these quantities reflect independent properties of the primordial field: 
the former characterizes the shape of the proto-halo region, while the latter is related to the number of particles included in the proto-halo region.

The second row presents heatmaps of Youden’s $J=\mathrm{TPR}-\mathrm{FPR}$ computed across the two-dimensional grid of thresholds in $\log_{10}(M_{\mathrm{sub}}/M_{\mathrm{lim}})$ and $c/a$. The value of the maximum $J$ is noted in the right bottom corner in each panel.
For $k_{\mathrm{fs}} = 10$ and $k_{\mathrm{fs}} = 12$, the statistic reaches $J = 1$, indicating nearly perfect discrimination between genuine and artificial subhaloes. 
Notably, the maximum of $J$ extends over a broad region in both $c/a$ and $M_{\mathrm{sub}}/M_{\mathrm{lim}}$ for $k_{\mathrm{fs}} = 10$, and primarily along $c/a$ for $k_{\mathrm{fs}} = 12$. 
This broad plateau arises because, when small-scale structures are strongly suppressed, only the main subhaloes are left, whose properties are highly distinct from those of artificial fragments, leading to clear separation in both mass and shape.
With increasing $k_{\mathrm{fs}}$, the distinction between genuine low-mass subhaloes and artificial fragments becomes less pronounced, leading to a decrease in the maximum value of $J$. 
This trend reflects the growing overlap between the two populations in the top panels.
Nevertheless, the classification performance remains relatively strong, with $J \simeq 0.7$ for $k_{\mathrm{fs}} = 20$, $30$, $35$, and $50$, 
and a slightly higher value of $J \simeq 0.8$ at $k_{\mathrm{fs}} = 24$.
Across different values of $k_{\mathrm{fs}}$, we find broadly consistent optimal thresholds in the $J$ statistic, 
with $\log_{10}(M_{\mathrm{sub}}/M_{\mathrm{lim}}) \simeq -1$ and $c/a \simeq 0.3$ for $k_{\mathrm{fs}}$ ranging from 20 to 50. 
For the strongly suppressed cases ($k_{\mathrm{fs}} = 10$ and $k_{\mathrm{fs}} = 12$), 
a single mass cut alone provides satisfactory discrimination between genuine and artificial subhaloes.

The bottom two rows display the true positive rate (TPR) and the false positive rate (FPR), with red crosses marking the locations of the corresponding maxima in the $J$ statistic. 
Allowing a more relaxed threshold for genuine subhalo identification in either $c/a$ or $M_{\mathrm{sub}}/M_{\mathrm{lim}}$ increases the TPR while simultaneously reducing the FPR.
For future applications with different scientific objectives, alternative operating points may be chosen to balance completeness (TPR) against contamination (FPR). 
We summarise three operating priorities per free–streaming wavenumber \(k_{\rm fs}\) in Table~\ref{tab:opt_thresholds_by_k}: 
(i) balanced: Maximize \(J\);
(ii) emphasis completeness (high TPR): within the near–optimal band \(J \ge J_{\max} - \varepsilon_J\) (with \(\varepsilon_J = 0.10\)), select the point with the highest TPR;
(iii) emphasis low contamination (low FPR): within the same band, select the point with the lowest FPR.
Thresholds are reported as $(\log_{10}(M_{\rm sub}/M_{\rm lim}),\,c/a)$.

\begin{table*}
  \centering
  \label{tab:opt_thresholds_by_k}
  \begin{tabular*}{\textwidth}{@{\extracolsep{\fill}}c|c|c|c|c|c|c|c|c|c|c}
    $k_{\rm fs}\,[h\,\mathrm{Mpc}^{-1}]$ & 10 & 12 & 20 & 24 & 30 & 35 & 50 \\
    \hline
    Max J & (-1.94, 0.47) & (-0.13, 0.10) & (-1.11, 0.35) & (-0.73, 0.25) & (-1.20, 0.20) & (-0.88, 0.27) & (-0.65, 0.05) \\
    High TPR & (-1.94, 0.47) & (-0.13, 0.10) & (-1.60, 0.17) & (-1.83, 0.25) & (-1.36, 0.20) & (-1.36, 0.27) & (-1.11, 0.25) \\
    Low FPR & (-1.94, 0.47) & (-0.13, 0.10) & (-1.11, 0.35) & (-0.73, 0.25) & (-1.20, 0.20) & (-0.88, 0.27) & (-0.65, 0.05) \\
  \end{tabular*}
  \caption{Optimal 2D thresholds for identifying artificial fragments using the combination of variables $\log_{10}(M_{\rm sub}/M_{\rm lim})$ and $c/a$, summarised per $k_{\rm fs}$. Each cell reports $\log_{10}(M_{\rm sub}/M_{\rm lim}),\,c/a)$. “Max~J” reports the global maximum of Youden’s $J$. “High~TPR” and “Low~FPR” are near–optimal alternatives chosen within the band $J \ge J_{\max}-0.10$, prioritising recall and precision, respectively. For comparison, \citet{Lovell2014} adopted $c/a = 0.16$ and $\log_{10}(M_{\rm sub}/M_{\rm lim})$ in the range $-0.40$ to $-0.22$. }
\end{table*}

\subsection{Identification of peak particles}
\label{subsec:self-similar}

\begin{figure*}
    \centering
    \includegraphics[width = 0.7\textwidth]{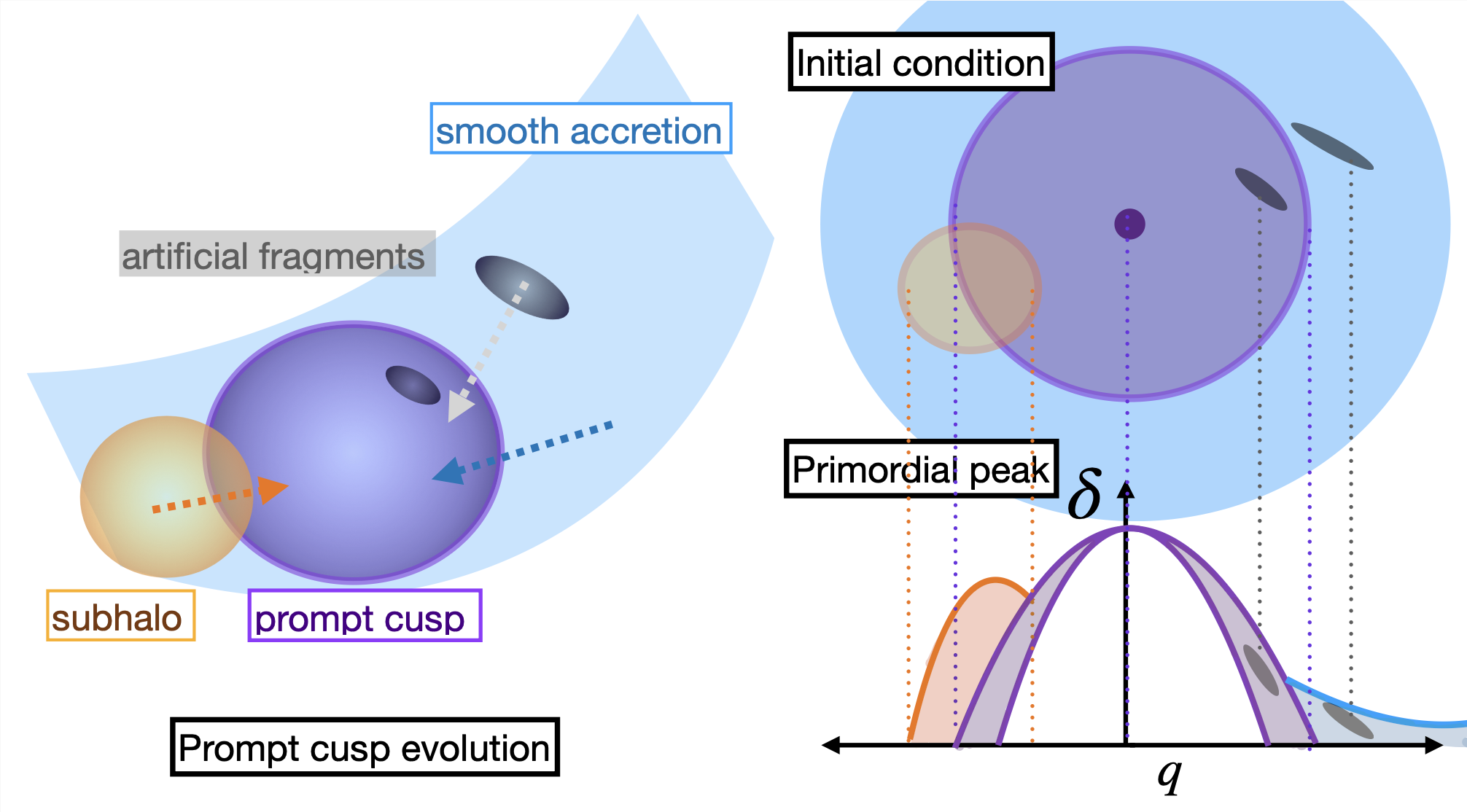}
    \caption{Schematic illustration of the three mechanisms breaking the self-similarity of prompt cusp formation (Sec.~\ref{subsec:self-similar}). The left panel shows the physical evolution of the prompt cusp (\textit{purple}) while the right panel shows the initial condition corresponding to this configuration. The upper right shows the spatial distribution of the main primordial peak (\textit{purple}) and environmental perturbations (\textit{orange}, \textit{blue} and \textit{grey}) on a projected coordinate plane, while the lower right displays the initial linear density field ($\delta$) as a function of Lagrangian distance from the centre ($q$) for these structures. The environmental perturbations manifest as:
    (1) Subhalo infall (\textit{orange}), arising from overlapping neighboring peaks in the initial conditions; (2) Smooth accretion (\textit{blue}), arising from diffuse background mass inflow; and (3) Artificial fragments (\textit{grey}), arising from numerical noise.
    }
    \label{fig:Abstract_IC}
\end{figure*}

\begin{figure}
    \centering
    \includegraphics[width=\columnwidth]{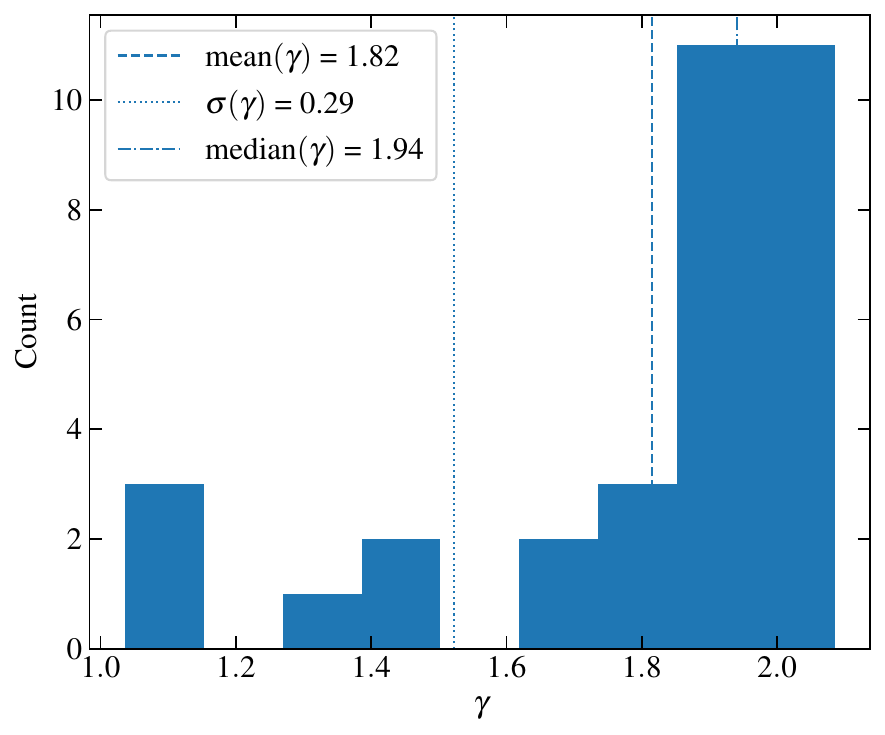}
\caption{Distribution of the fitted peak shape index $\gamma$. The histogram aggregates all haloes and all free–streaming realisations included in the analysis, with the summary statistics are annotated in the panel. The mean, standard deviation and median are 1.77, 0.31 and 1.91, respectively.
    }
    \label{fig:Gamma}
\end{figure}

\begin{figure*}
    \hspace*{-1.8cm}
    \centering
    \includegraphics[width = 1.2\textwidth]{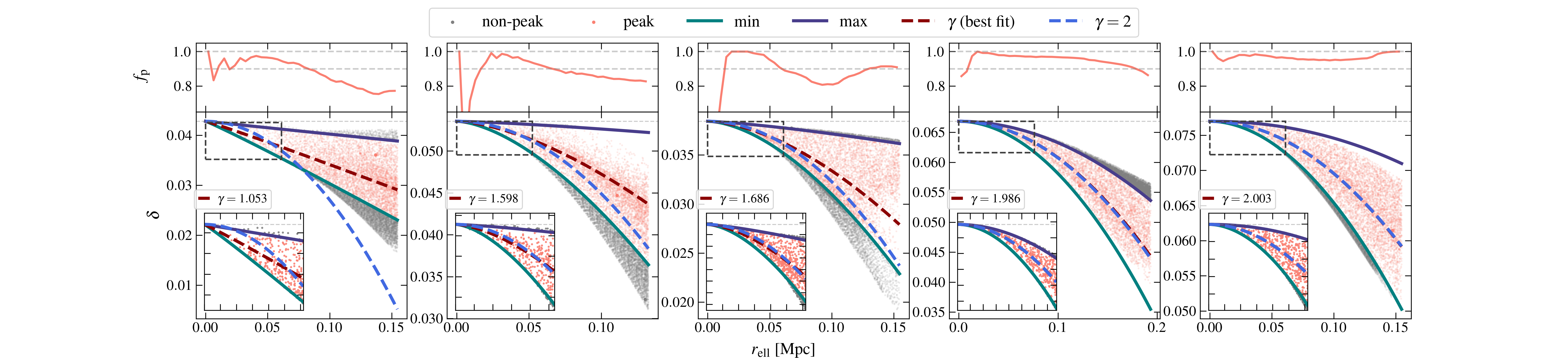}
    \caption{Diagnostic of the peak-selection and profile-fitting procedure described in Sec.~\ref{subsec:self-similar}.
    Each panel corresponds to a different peak selected from one $\gamma$-bin in Figure~\ref{fig:Gamma}. From left to right the Halo ID = $\{150, 150, 260, 108, 108\}$ and $k_{\rm sf}$ are $=\{30, 35, 30, 24, 30\}\,h\,\mathrm{Mpc}^{-1}$ we find $\gamma=\{1.052,\,1.598,\,1.686,\,1.986,\,2.003\}$.
    Points show the linear overdensity $\delta$ of all particles in the peak neighbourhood as a function of their elliptical radius $r_{\rm ell}$: non-peak particles are plotted in grey, and peak particles - defined by the envelope criterion in Sec.~\ref{subsec:self-similar} - are plotted in salmon.
    The teal and dark purple curves denote the minimum and maximum envelope fits, respectively.
    The dark red dashed curve shows the best-fit free-index $\gamma$ profile while the blue dashed curve shows the comparison with fixed $\gamma=2$.
    The fitted $\gamma$ for each peak is reported in the legend of the corresponding panel.
    A dashed grey rectangle on each main panel marks the exact extent of the central region used for the fits, with its zoom-in view in the sub-panel. A grey dashed line in the main panel shows the maximum $\delta$ at the $r_{\rm ell} = 0$.
    Above each main panel, a subpanel shows the fractions of peak particles as functions of \(r_{\rm ell}\), providing a quantitative view of the central dominance of peak material and the progressive increase of non-peak material with increasing radius. The horizontal dashed lines mark the fraction of 1 and 0.9.
    }
    \label{fig:fitting_peak}
\end{figure*}


The analytic model of prompt cusp formation developed by \citet{White2022} assumes a self-similar collapse of concentric mass shells around a primordial density peak. Under this assumption, the initial density field near the peak can be approximated by a power-law profile in Lagrangian space:
\begin{equation}
\label{eq:ss_ic}
    \rho(q, t) = \bar\rho(t)[1 + \delta_{\rm p, max}(1 - 1/6(q/r_{\rm p})^\gamma)],
\end{equation}
where $q$ is the Lagrangian radial distance from the peak centre, $\delta_{\rm p, max}$ is the peak overdensity, and $\gamma$ is the power-law index controlling the steepness of the profile. This generalisation replaces the specific second-order expansion used in \citet{White2022} and allows more flexibility in fitting realistic peak shapes by effectively incorporating information from the higher order terms while maintaining the self-similarity condition.

While this self-similar framework provides a tractable analytic description of early halo formation, it may be violated under realistic cosmological conditions. We identify three mechanisms that can break self-similarity and the resultant phenomena, also illustrated in Fig.~\ref{fig:Abstract_IC}:
\begin{enumerate}
    \item {\it Peak overlap - subhalo infall}: In a cosmological context, neighbouring peaks in the initial density field may partially overlap. If one peak collapses \textit{before} its material is accreted by the main peak, the resulting structure appears as a distinct halo that subsequently merges with the main halo. 
    \item {\it Mass inflow - smooth accretion}: Material originating outside the progenitor peak - either from neighbouring peaks that have not yet collapsed or from the diffuse background - is accreted directly onto the main halo prior to the neighbours’ collapse. In this case, no merger occurs, but an elevated accretion rate may indicate that the main halo is incorporating matter not originating from its own progenitor peak.
    \item {\it Artificial fragments}: Spurious haloes may form both within and outside the boundaries of the primordial peak, disrupting the self-similar evolution of the mass shells. Artificial fragments originating outside the peak are typically accreted by the main halo at later times, producing disturbances analogous to {\it subhalo infall} events \citep{Ondaro2024}. Conversely, those forming within the peak are incorporated prior to the formation of the prompt cusp, preventing the emergence of a coherent power-law density profile.
\end{enumerate}
In Fig.~\ref{fig:Abstract_IC}, the right upper panels depict the initial conditions in Lagrangian space, where the central primordial peak (\textit{purple}) is surrounded by specific environmental features corresponding to the three identified mechanisms: an overlapping neighboring peak (\textit{orange}), the diffuse background field (\textit{blue}), and small-scale discreteness noise (\textit{grey}). Their contribution to the overdensity field is illustrated in the lower right panel, with dotted lines linking the corresponding projected structures with the spatial extent on the Lagrangian coordinate $q$. The left panel shows the corresponding physical evolution in Eulerian space. Here, the neighboring peak manifests as a distinct subhalo merging with the main structure (subhalo infall), while the diffuse background contributes to a steady mass inflow (smooth accretion). The small-scale fluctuations appear as dense artificial fragments that fall into the potential well, distinguishing them from the smooth background component. 

All three processes - {\it subhalo infall}, {\it smooth accretion}, and the {\it formation of artificial fragments} - introduce anisotropy into the halo. However, only the first and third mechanisms directly disrupt the coherent, self-similar movement of mass shells within the peak. These disruptions are more likely to significantly alter the internal density structure. The material from outside of the peak that falls in smoothly will slowly accumulate at the outskirts of the halo, not affecting the density distribution in the centre. Nevertheless, if the total enclosed mass $M(<r_{\rm M})$ in the equation of motion (Eq.~\ref{eq:eom}) remains dominated by material originating from the peak - where the density is highest - the resulting gravitational potential is only mildly perturbed, allowing the central structure (i.e., the prompt cusp) to remain largely intact.

Given that the central gravitational potential remains dominated by peak material despite external perturbations, it is feasible to recover the prompt cusp at later times by separating self-similarly collapsed particles from those accreted later. 
Our method of separating the peak particles from the rest of the particles is detailed below:
\begin{enumerate}
    \item We first identify local maxima of the linearly extrapolated density field within the Lagrangian region of the target halo, and define a neighbourhood of peak scale $r_{\rm p}$ as done in Sec.~\ref{subsec:spurious}. We select all halo particles whose Lagrangian coordinates lie within the neighbors of the peaks.
    \item For the particles in this neighbourhood we measure the overdensity $\delta$ as a function of their distance to the peak centre. To capture the intrinsic asphericity of realistic peaks, we shift all particle positions into a peak-centred frame and fit a quadratic function. $\delta(\mathbf{q}) \simeq \delta_0 + \mathbf{q}^{\mathsf T}\mathbf{Q}\mathbf{q}$ where $\mathbf{q}$ is the 3D Lagrangian position relative to the peak centre and $\mathbf{Q}$ is a symmetric $3\times3$ matrix constrained such that the spherical limit corresponds to $Q_{xx}=Q_{yy}=Q_{zz}=Q$ and off-diagonal terms vanish.
    \item We then define the ellipsoidal radius \(r_{\rm ell}\) by evaluating the quadratic form at particle positions (with respect to the peak centre) and normalising by the local quadratic scale. Numerically this is done as
    $r_{\rm ell}(\mathbf{q}) \equiv \left[ \mathbf{q}^{\mathsf T}\mathbf{Q}\mathbf{q}/Q \right]^{1/2}$, which maps nested isodensity shells into $r_{\rm ell}=\mathrm{const}$ surfaces. 
    \item We then construct a profile $\delta(r_{\rm ell})$ (shown as grey and salmon points in Figure~\ref{fig:fitting_peak}) by binning particles in $r_{\rm ell}$ and, in each of the 50 radial bins, recording both the maximum and minimum values of $\delta$ as running quantiles, $
    \delta_{\rm min}(r_{\rm ell}) \equiv Q_{q_{\rm min}}\{\delta\,|\,r_{\rm ell}\}, \delta_{\rm max}(r_{\rm ell}) \equiv Q_{q_{\rm max}}\{\delta\,|\,r_{\rm ell}\},$
    with $(q_{\rm min},q_{\rm max})=(0.01,0.99)$. We fit the two envelopes with
    $\delta_{[m]}(r_{\rm ell}) \simeq \delta_{0} + \alpha{[m]} r_{\rm ell}^{\gamma[m]}$, where $[m]$ represents both the maxima and the minima and $\alpha$, $\gamma[m]$ are the fitting parameters and $\delta_0$ is the maximum value of $\delta$. Fixing $\delta_0$ ensures that the centre of the peak is included. The envelopes are estimated using only particles within a central zoom domain, \(r_{\rm ell}\le 0.4\,r_{\rm p}\) to avoid bias introduced by other overlapping peaks at the boundary.
    \item With the profile defined as above, we can classify \textit{all} particles belonging to the main peak. We evaluate the corresponding envelope profile $\delta_{\rm max}(r_{\rm ell})$ and $\delta_{\rm min}(r_{\rm ell})$ from the upper and lower envelope fit (shown as dark purple and teal lines in Figure~\ref{fig:fitting_peak}), and compare it to the particle’s own linear overdensity $\delta$. Particles whose overdensity lies between the fitted peak envelopes,
    $\delta_{\rm min}(r_{\rm ell})\le\delta(r_{\rm ell}) \le \delta_{\rm max}(r_{\rm ell}),$
    are tagged as `peak particles' and the rest as `non-peak' particles. They are shown in salmon and grey in Figure~\ref{fig:fitting_peak}, respectively. 
    \item We estimate the logarithmic slope of the peak profile by the fitting the $\delta(r_{\rm ell}) \simeq \delta_{0} + \alpha_{\gamma} r_{\rm ell}^\gamma$ profile using the peak particles (shown as dark red lines in Figure~\ref{fig:fitting_peak}); the resulting best-fitting slope, which we denote $\gamma$, is used as our estimator of the peak-shape index and we identify in the self-similar model of Eq.~\ref{eq:ss_ic} and Eq.~\ref{eq:pw_density}. We also fix $\gamma = 2$ and show the fitting results in Figure~\ref{fig:fitting_peak} as dark blue as comparison.
    
\end{enumerate}

The rationale for selecting particles below the fitted maximum envelope is that within the halo’s Lagrangian region, particles whose $\delta$ exceed this envelope at a given $r_{\rm ell}$ are typically associated with sharper, secondary maxima superposed on top of the main peak that do not follow the trend of the central $\delta(r_{\rm ell})$ profile. By contrast, particles with $\delta$ below the minimum envelope are treated as noise and are excluded, although this constitutes a very small fraction because our near-peak selection has linear overdensity dominating the noise. In other words, the peak-identification scheme is valid as a criterion for selecting material that is consistent with the smooth power-law density distribution around the primordial peak profile, and thus consistent with the self-similar assumption. Combining the information on the overdensity distribution of environmental perturbations in Fig.~\ref{fig:Abstract_IC}, this confirms that our procedure of selecting the peak-only profile is a valid method for distinguishing the main peak from environmental influences. An exception exists for cases where artificial fragments form inside the prompt cusp region, as they are indistinguishable using the envelope cuts.

The histogram in Figure~\ref{fig:Gamma} summarises the $\gamma$ values in the 64 simulations, with the mean, standard deviation and median of the distribution being 1.77, 0.31 and 1.91, respectively. 
These results show that despite $\gamma$ has majority distribution close to 2, it spans a substantial range, therefore the theoretical prompt-cusp slope $\lambda(\gamma)=-6\gamma/(3+2\gamma)$ can deviate appreciably from the value previously used in the self-similar model; for example, $\gamma\simeq 1$ implies $\lambda\simeq -6/5 \approx -1.2$. 

While the largest peak in each region typically yields a value of \(\gamma\) very close to 2, the full population of identified peaks in the halo Lagrangian region exhibits a wider spread, including values close to \(\gamma \simeq 1\). This occurs because the progenitor peak used to define the prompt cusp - selected as the peak with the highest mean density in the innermost five logarithmic bins above $r_{\rm conv}$ - does not always coincide with the largest peak in the Lagrangian region. In particular, during major merger phases the rapidly evolving internal dynamics can cause the instantaneous density centre to be associated with a smaller, originally secondary primordial peak whose material temporarily dominates the inner density, leading to inferred \(\gamma\) values close to 1 for the ostensibly 'main' peak. This phenomenon highlights a general challenge in defining halo centres in hierarchically growing systems and emphasises that detailed inspection of the merger state is needed to mitigate such mis-association, for example by restricting profile measurements to snapshots in which the halo is dynamically relaxed. To address this in our analysis, we focus on early snapshots immediately following prompt cusp formation, thereby minimizing the probability that the halo is undergoing a major merger.

In Figure~\ref{fig:fitting_peak} we show five examples of the peak-particle identification scheme for one peak in each $\gamma$-bin in Figure~\ref{fig:Gamma}. The best-fit power-law indices are close to 2 across the sequence; from left to right the Halo ID $150, 260, 150, 108, 108$ and $k_{\rm sf}$ are $=\{30, 30, 35, 24, 30\}\,h\,\mathrm{Mpc}^{-1}$ we find $\gamma=\{1.052,\,1.686,\,1.598,\,1.986,\,2.003\}$. Within the central region, the highest-overdensity particles are consistently classified as peak particles, confirming that the enclosed mass $M(<r_{\rm M})$ in Eq.~\ref{eq:eom} is dominated by peak material at least at the initial condition. The fraction of peak particles within the central $0.4r_{\rm p}$ region across the five simulations is $\{0.923, 0.951, 0.957, 0.974, 0.972\}$ whereas over the entire peak it is $\{0.811, 0.872, 0.859, 0.932, 0.970\}$. The result of this fraction as functions of \(r_{\rm ell}\), is shown in the upper panels.
Propagating forward in time, this dominance implies that the motion of the central shells is well described by Eq.~\ref{eq:eom}, and the peak particles trace the resulting power-law density profile of the halo, consistent with self-similar evolution.

To describe the internal matter distribution inside the halo, we compute the spherically averaged density profile between the convergence radius and the virial radius.
The convergence radius, $r_{\rm conv}$, defines the innermost radius at which the density profile is numerically reliable, and is calculated using the criterion from \citet{Power2003}:
\begin{equation}
    0.6 = \frac{\sqrt{200}}{8}\frac{N(r_{\rm conv})}{{\rm ln}N(r_{\rm conv})}\left(\frac{\overline{\rho}(r_{\rm conv})}{\rho_{\rm crit}}\right)^{-\frac{1}{2}}
\end{equation}
where $N(r_{\rm conv})$ is the number of particles within $r_{\rm conv}$, $\overline{\rho}$ is the mean enclosed density, and $\rho_{\rm crit}$ is the critical density of the Universe.
The density profile is computed in $50$ logarithmic bins between $r_{\rm conv}$ and $r_{\rm 200}$. 
To mitigate the impact of inhomogeneities caused by substructures, we compute the profile by dividing the solid angle into 100 angular bins evenly spaced in azimuth and polar coordinates, and taking the median of the radial (line-of-sight) bins. 

To estimate the spatial extent of the size of the halo, we define the caustic radius, following the method of \citet{Diemer2014}. 
The caustic radius ($r_{\rm c}$) is defined as the radius at which the logarithmic slope ($\mathrm{d}\ln\rho/\mathrm{d}\ln r$) - calculated from $\ln(\rho)$ smoothed with a Savitzky-Golay filter (window length 15 bins, polynomial order 4) - attains its first minimum.
We compute the density profile in physical coordinates, rather than comoving coordinates, because the halo behaves as a gravitationally bound structure decoupled from the background Hubble flow, enabling direct comparison with analytical collapse models and with other simulated haloes at different redshifts.

By tracking the particle ID of the selected peak particles at later times, we use this subset of particles to construct a separate density profile called the `peak profile', which isolates the present-day contribution of matter originating in the primordial peak from the rest of the halo. When multiple progenitor peaks are associated with the main subhalo, we compute a peak profile for each candidate and adopt as the representative one the profile whose mean density in the innermost five logarithmic bins immediately above the convergence radius $r_{\rm conv}$ is highest; this criterion selects the progenitor whose material dominates the central mass budget.

Following the analytical calculation in \cite{White2022}, the density profile of the prompt cusps can be described as
\begin{equation}
\label{eq:pw_density}
    \rho(r) = Ar^{-(6\gamma)/(3+2\gamma)},
\end{equation}
where $r$ is the Eulerian radial distance and $A$ is the amplitude of the density profile. For $\gamma = 2$, this recovers the canonical prompt cusp slope of $-12/7$ in \cite{White2022}. 
Following \citet{White2022}, we compute the cusp amplitude \(A\) by numerically integrating the similarity equations and matching at second apocentre. 

\subsection{Description of the structure evolution}
\label{subsec:density_profile}

To characterise the evolution of the inner density structure, we fit four model families to the spherically averaged density profiles of both haloes and peak particles: three power–law models with slopes $\lambda = -3/2$, $\lambda = -12/7$, and a free slope $\lambda(\gamma)$, and an NFW profile. For brevity, we refer to these as the $3/2$ profile, the $12/7$ profile, the $\lambda$ profile, and the NFW profile, respectively.

The density profiles of haloes and peak particles are fit with each of these model families, and the goodness of fit is quantified using the reduced $\chi^2_\mu$ statistic. To compare models with different numbers of free parameters, we use the small–sample corrected Akaike Information Criterion, $\mathrm{AIC}_\mathrm{c}$; among the models considered, the one with the lowest $\mathrm{AIC}_\mathrm{c}$ is preferred. The details of the fitting procedure and the computation of $\mathrm{AIC}_\mathrm{c}$ are provided in Appendix~\ref{app:profile_fitting}.

\section{Results}

\begin{table*}
\centering
\begin{tabular*}{\textwidth}{@{\extracolsep{\fill}}c|c|c|c|c|c|c|c|c}
\hline\hline
$k_{\rm fs}[h\,{\rm Mpc}^{-1}]$ & 7 & 10 & 12 & 20 & 24 & 30 & 35 & 50 \\
\hline
Halo 108& [ ] & + & +& +& +& +& +& +\\
Halo 150& [ ] & [ ]& +& +& +& +& +& +\\
Halo 206& [ ] & [ ]& =& =& =& =& +& +\\
Halo 225& [ ] & [ ]& [ ]& =& =& =& +& +\\
Halo 260& [ ] & [ ]& [ ]& =& =& =& +& +\\
Halo 305& [ ] & [ ]& [ ]& =& =& =& =& +\\
Halo 337& [ ] & [ ]& [ ]& =& =& =& =& +\\
Halo 443& [ ] & = & =& +& +& +& +& +\\
\hline\hline
\end{tabular*}
\caption{Summary of halo formation states and contamination from artificial fragmentation across different free-streaming cut-off scales. Due to the strong small-scale suppression, simulations where haloes fail to form by $z = 0$ are marked with “[ ]” (18 simulations, 28\%). Simulations in which early artificial fragmentation is visually identified during prompt cusp formation are marked with “=” (20 simulations, 31\%). Halo formed without significant interaction with artificial fragments are marked with "+" (26 simulations, 41\%). haloes are ordered by decreasing mass from top to bottom, and $k_{\rm fs}$ values increase left to right.}
\label{tab:Halo_formation}
\end{table*}
We inspect the evolution of all 64 simulations and classify the overall outcome of each run into three categories, as summarised in Table.~\ref{tab:Halo_formation}. 
First, due to the strong small-scale suppression, simulations in which no bound halo forms by $z = 0$ are marked with ``[ ]''; this category contains 18 out of 64 simulations, corresponding to 28\% of the sample. 
Second, simulations in which early artificial fragmentation is clearly visible during the prompt–cusp phase are marked with ``=''; this group comprises 20 simulations, or 31\% of the total. 
Finally, simulations in which the main halo forms without significant interaction with artificial fragments are marked with ``+''; this class includes the remaining 26 simulations, accounting for 41\% of the sample.
Fewer haloes form at low $k_{\rm fs}$ (large-scale cutoffs), with no formation observed at $k_{\rm fs} = 7\,h\,{\rm Mpc}^{-1}$. All haloes form at $k_{\rm fs} = 50\,h\,{\rm Mpc}^{-1}$. This confirms that our $k_{\rm fs}$ range spans the threshold of halo formation. 

Artificial fragmentation is also more prominent at low $k_{\rm fs}$, particularly for low-mass haloes near the $\Delta^2$ peak. For example, haloes 206, 225, and 260 remain affected by artefacts up to $k_{\rm fs} = 30,h,{\rm Mpc}^{-1}$, whereas haloes 305 and 337 are only clean at the highest $k_{\rm fs}$. An exception is Halo 443, which shows reduced contamination, possibly due to its higher effective resolution.

\subsection{Examples of different evolutionary pathways}
\label{subsec:example}
\begin{figure*}
    \centering
    \includegraphics[width = \textwidth]{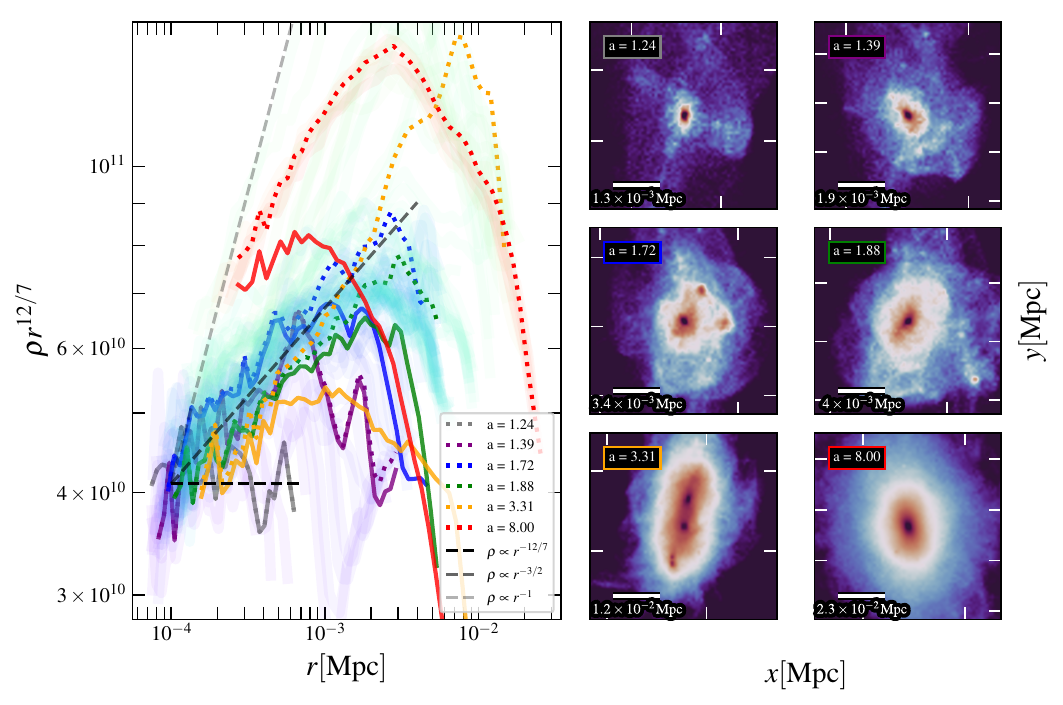}
    \caption{Time evolution of the density profiles versus physical radius $r$ (\textit{left}) and corresponding projected density maps (\textit{right}) for selected haloes. 
    Profiles are scaled by $r^{12/7}$, such that a horizontal line (dashed black line) corresponds to the theoretical value predicted by self-similar collapse model in \citep{White2022}, while the prompt cusp slope of $-3/2$ \citep{Delos2023} is indicated by the dashed dark grey line. The dashed light grey line dashed line marks a slope of $-1$, corresponding to the central asymptotic behaviour of the NFW profile.
    Line colours indicate simulation snapshots at key evolutionary stages: \textit{grey} marks the initial collapse of the peak, \textit{purple} corresponds to the formation of the prompt cusp, \textit{blue} captures the onset of a minor merger, \textit{green} follows the post-merger state, \textit{orange} represents an ongoing major merger, and \textit{red} indicates the post-major merger configuration. These stages are also shown in the legend.
    The simulation times are expressed in terms of the rescaled scale factor $a = a_s/a_{\rm c}$, where $a_s$ is the scale factor at each snapshot and $a_{\rm c}$ is the scale factor at which the peak undergoes collapse. Dotted lines denote the density profile of all halo particles, while solid lines show the profile computed from peak particles only. The lighter coloured lines in the background are the density profiles at all snapshots, plotted with the same colour scale as the thicker lines.
    }
    \label{fig:Evo_Profile}
\end{figure*}
In this section, we examine the evolution of the density profiles of individual haloes, complemented by visual inspection of their corresponding density maps. Based on the variation in these profiles and their agreement with structural features in the maps, we identify six representative evolutionary stages. We focus on the time evolution of both the total and peak-only density profiles, assessing their consistency and interpreting the divergence in the context of the physical mechanisms outlined in Section~\ref{subsec:self-similar}.

Figure~\ref{fig:Evo_Profile} presents the time evolution of the spherically averaged density profiles (\textit{left}) and the projected density maps (\textit{right}) for a representative halo. Profiles are scaled by $r^{12/7}$, such that a horizontal line corresponds to the theoretical prompt cusp slope of $-12/7$ \citep{White2022}. Density slopes are shown using dashed lines (see caption for conventions).
The profiles are colour-coded according to the rescaled simulation time, $a = a_{\rm s}/a_{\rm c}$, where $a_{\rm s}$ is the simulation scale factor and $a_{\rm c}$ is the scale factor at peak collapse, calculated as:
\[
a_{\rm c} \;=\; \frac{1.686\,a_{\rm ini}}{\delta_{\rm peak}}, \qquad a_{\rm ini}=\frac{1}{1+z_{\rm ini}},
\] 
progressing from grey and purple (early times), through blue, green, and orange, to red (late times), as labelled in the right-hand legend. This removes trivial phase shifts associated with different peak heights and enables fair comparisons of prompt-cusp onset and persistence.

Dotted lines represent the density profile of all halo particles, while solid lines correspond to the subset of `peak' particles associated with the primordial overdensity, as identified in Section~\ref{subsec:self-similar}. 

We now describe the six stages in detail, based on their appearance in the rescaled density profiles and projected maps.
\begin{itemize}
    \item $a = 1.24$: The grey line marks the initial collapse of the peak - before it reaches a stable profile - where the density profile briefly appears as a power law with a slope of $-12/7$. 
    The overlap of peak (grey solid) and total (grey dashed) profiles reflects that all particles originate from the initial peak.
    As this is a transient configuration, the apparent agreement with the analytical slope of $-12/7$ does not contradict prior simulations that report a stable slope of $-3/2$ for fully formed prompt cusps.
    \item $a = 1.39$: The purple line corresponds to the formation of the prompt cusp. At this stage, the density profile settles into a stable power-law form, with a slope of approximately $-3/2$ - in agreement with previous simulation results (e.g. \citealt{2013Anderhalden, 2018Delos, 2023Delos, Delos2023}). A slight deviation between the total (\textit{dashed}) and peak-only (\textit{solid}) profiles emerges at larger radii, due to the onset of smooth accretion from surrounding material. However, the excellent agreement at small radii indicates that peak particles continue to dominate the central gravitational potential, rendering the accreted outer material dynamically subdominant within the cusp region.
    \item $a = 1.72$: The blue line captures the onset of a minor merger, during which two small over-densities become visible near the halo centre. This event leads to a noticeable rise in the total density profile at large radii compared to the earlier $a = 1.39$ snapshot. A modest increase in central density is also observed, likely due to a transient interaction with a nearby filament. As will be discussed later in Section~\ref{sec:example_lls}, this filamentary structure sweeps across the halo, briefly enhancing the central density and potentially delivering the two additional substructures. Although some material is retained post-interaction, the central density declines again once the halo detaches from the filament. The presence of infalling material is reflected in the growing discrepancy between the total density profile (blue dashed) and the peak-only profile (blue solid), particularly in the outskirts.
    \item $a = 1.88$: The green line corresponds to the post-merger state. Following the transient central enhancement during the minor merger, the central density relaxes and once again aligns with the pre-merger profile, following the $r^{-3/2}$ power-law characteristic of the prompt cusp. However, the outskirts do not fully recover - a substantial discrepancy remains between the total density profile (green dashed) and the peak-only profile (green solid). This indicates that the outer regions are still dominated by material accreted during the earlier interaction, whether from the minor merger itself or accompanying inflow along the local filament.
    \item $a = 3.31$: The orange line represents an ongoing major merger. During this phase, the density profile is heavily disrupted due to the strong gravitational perturbation introduced by an infalling halo of comparable mass, which is visible near the centre of the system. This interaction distorts the central structure of the original halo, as evidenced by the significant deviation and flattening in the peak-only density profile (orange solid line). Simultaneously, the total density profile (orange dashed) shows a substantial rise in the outskirts, reflecting the accumulation of material from the merging companion and the overall disturbance of the halo’s equilibrium configuration.
    \item $a = 8.00$: The red line indicates the post-major merger configuration at redshift $z = 0$. At this stage, the halo has fully transitioned into an NFW-like structure, with the total density profile exhibiting an inner slope close to $-1$, characteristic of the NFW profile. Both the total (red dotted) and peak-only (red solid) density profiles have deviated from the prompt cusp form, indicating that the original self-similar structure has been destroyed by the major merger. Although the two profiles still appear to converge in the innermost region, the limited resolution of the simulation prevents a definitive assessment of the central structure. Nevertheless, it is evident that the peak material continues to occupy the centre of the halo.
\end{itemize}

This sequence of evolutionary stages highlights the distinct roles of different physical mechanisms in shaping the central density structure of haloes. The initial prompt cusp, formed via self-similar collapse, is shown to be resilient to early smooth accretion and minor mergers - both of which primarily contribute mass to the outer regions without significantly perturbing the central potential. However, major mergers, involving haloes of comparable mass, generate strong dynamical disturbances that disrupt the coherent motion of peak material, destroying the central cusp and transforming the total density profile into the NFW form.

While Figure~\ref{fig:Evo_Profile} illustrates how the specific events affect the total and peak density profiles of an individual halo, the full set of simulations reveals that haloes typically evolve through one or more of three distinct interaction modes: (i) mergers with spurious substructures (artificial fragmentation), (ii) interactions with nearby structures such as filaments, fly-by haloes and minor mergers, and (iii) major mergers with other comparable-mass haloes.
These pathways are summarised schematically in Figure~\ref{fig:Abstract}. 

The cartoon follows the evolution of the halo from left to right in time, beginning with a centrally concentrated prompt cusp (purple) and ending with a fully developed NFW halo (red). Density profiles in each panel are scaled by $r^{12/7}$, following the same convention as Figure~\ref{fig:Evo_Profile}, so that a flat dashed black line corresponds to the $-12/7$ prompt-cusp slope. Dashed dark grey and light grey lines indicate reference slopes of $-3/2$ (analytical self-similar prediction) and $-1$ (NFW), respectively. 

The three pathways are not mutually exclusive and may occur in combination or repetition, but are separated here to illustrate their distinct signatures. 
In the artificial fragmentation pathway (\textit{blue}), early interactions with artificial fragments can deposit dense clumps into the halo interior. This significantly elevates the density at all radii and disturbs the original power-law structure, as reflected by the upward shift and flattening of the schematic profile. Such interactions can prematurely disrupt the prompt cusp.
In the local structure disturbance pathway (\textit{green}), the halo encounters nearby structures, such as filaments, fly-by halo, and experiences minor merger. These interactions induce an increase in mass accretion, and result in a modest changes in the outer profile while largely preserving the inner power-law slope. The schematic shows slight buildup at large radii, representing smooth or non-disruptive accretion.
The major merger pathway (\textit{yellow}) involves merger with a massive halo. The central density profile becomes sharply distorted, and the power-law shape is lost. The schematic reflects this transition with a curve that approach NFW at the centre, with additional material at the outskirts, capturing the non-equilibrium structure during the merger event. 

Arrow style encodes origin: solid arrows denote physical processes (major merger with halos and disturbance from local structures), while dotted arrows denote numerical artefacts (merger with artificial fragments).
To aid interpretation, the schematic in Figure~\ref{fig:Abstract} arranges the evolutionary stages from left to right in order of when each process typically occurs during structure formation: spurious mergers occur earliest, followed by local interactions, and finally major mergers. The vertical shaded regions in the background represent the relative mass or size of the halo at each stage, which grows with time. 

Detailed examples of each pathway are presented in the next subsections: Section~\ref{sec:example_merger} demonstrates the consequences of a major merger, Section~\ref{sec:example_lls} presents an interaction between the halo and a filament, and Section~\ref{sec:example_spurious} shows a case of merging with artificial fragments just after the formation of the prompt cusp.

\begin{figure*}
    \centering
    \includegraphics[width = \textwidth]{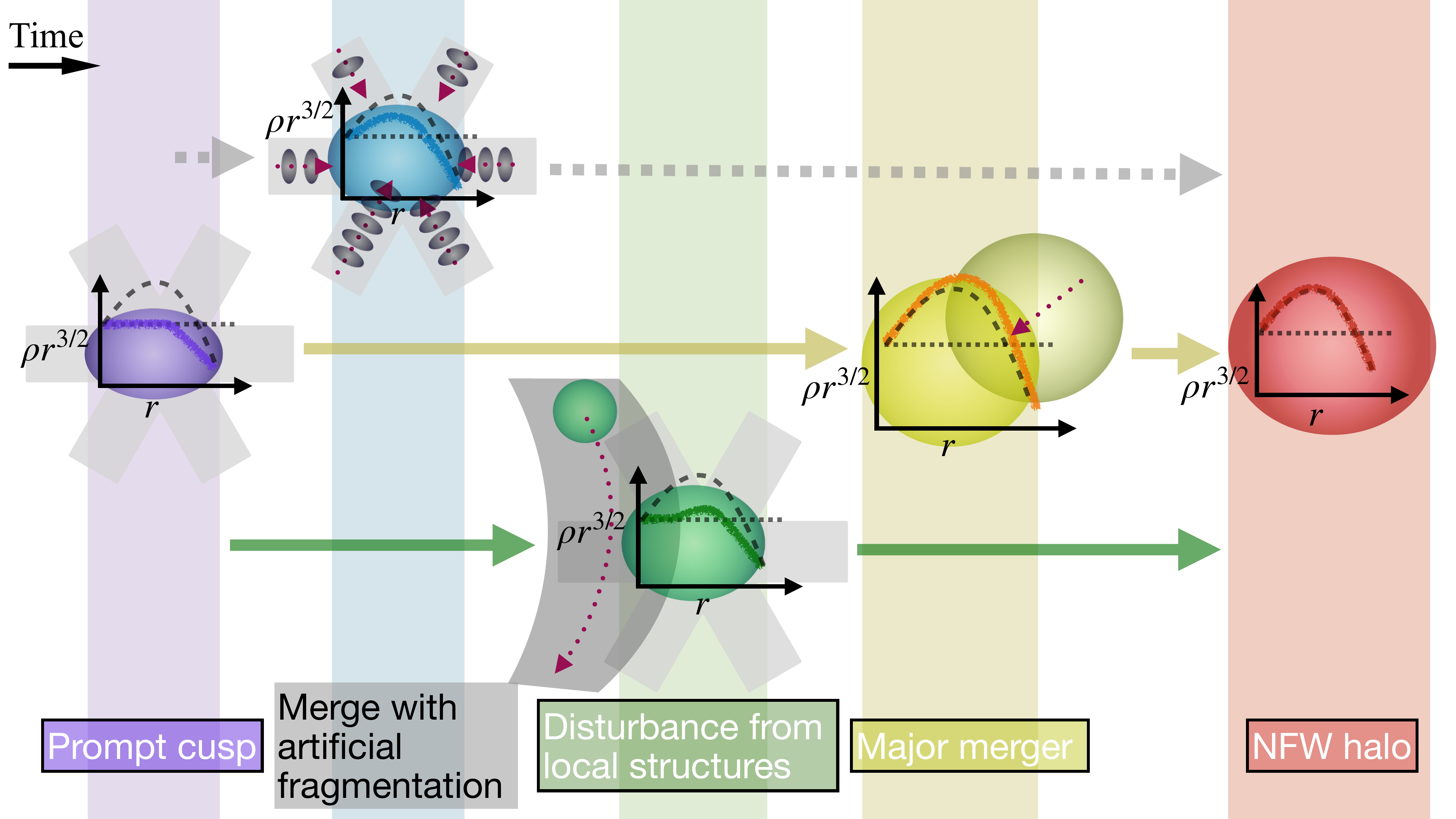}
    
    \caption{
    Cartoon illustration of three evolutionary pathways from prompt cusp formation to an NFW halo. Time progresses left to right. Each panel shows the halo structure and corresponding density profile (coloured line), scaled by $r^{3/2}$. Horizontal dotted and dashed lines represent the $r^{-3/2}$ and NFW profile, respectively. Time flows from left to right; background shading represents increasing halo mass.
    Pathways include mergers with artificial fragments (blue), interactions with local structures (\textit{green}), and major mergers (\textit{yellow}). These processes progressively alter the initial prompt cusp profile (\textit{purple}) into an NFW-like form (\textit{red}), with varying impact on the central density. Examples are shown in Figures ~\ref{fig:Evo_Merge},  ~\ref{fig:Evo_Filament}, and ~\ref{fig:Evo_Spurious}.
    The horizontal layout encodes typical timing: prompt-cusp destruction by artificial fragments (\textit{blue}) tends to occur early, whereas major mergers (\textit{yellow}) typically occur later. Arrow style encodes origin: solid arrows denote physical processes, while dotted arrows denote numerical artefacts. The same distinction is reflected in the styling of the accompanying text boxes.
    }
    \label{fig:Abstract}
\end{figure*}

\subsubsection{Major mergers}
\label{sec:example_merger}
\begin{figure*}
    \centering
    \includegraphics[height = 0.9\textheight]{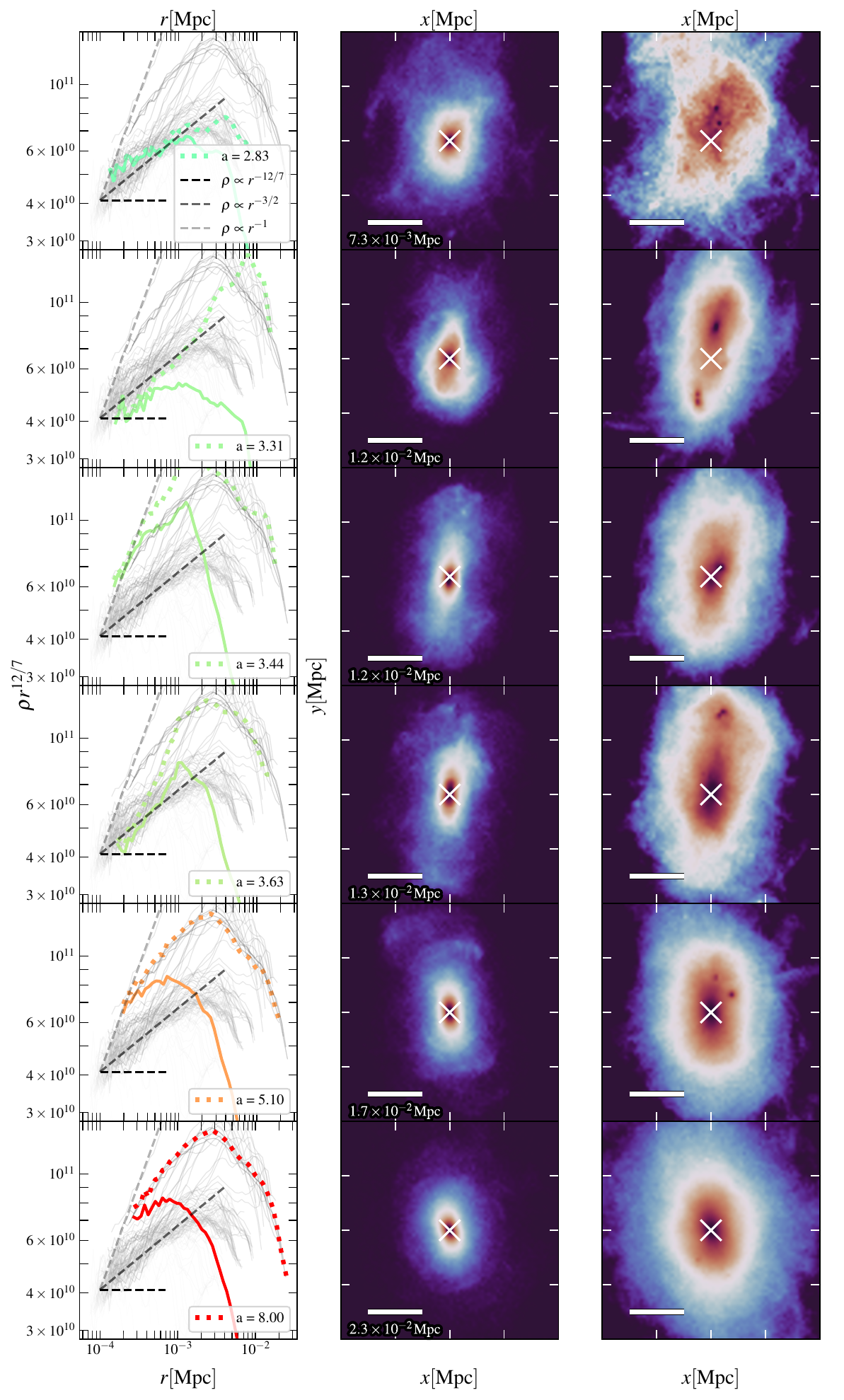}
    \caption{Evolution of the configuration and density profile versus physical radius $r$ of a halo undergoing a major merger. Snapshots are arranged in time from \textit{top} (before the merger) to \textit{bottom} (after the merger). Each row corresponds to a snapshot, showing (\textit{left}) the density profiles, (\textit{middle}) the spatial distribution of peak particles, and (\textit{right}) the distribution of non-peak particles. In the left panels, solid lines represent the density of peak particles, while dotted lines show the total density including both peak and non-peak components. Line colours indicate simulation snapshots. Line colours indicate different simulation snapshots. As in Figure~\ref{fig:Evo_Profile}, the profiles are scaled by $r^{12/7}$, such that a horizontal black dashed line corresponds to the theoretical slope from \citet{White2022}. For reference, slopes of $-3/2$ and $-1$ are shown by the dark-grey and light-grey dashed lines, respectively.
    The white cross in each density map marks the main subhalo centre as identified by \texttt{SUBFIND}.}
    \label{fig:Evo_Merge}
\end{figure*}

Figure~\ref{fig:Evo_Merge} illustrates the evolution of a halo undergoing a major merger, arranged in columns from top to bottom in increasing simulation time (i.e. early to late stages). Each row corresponds to a snapshot, showing (left) the spherically averaged density profile, (middle) the spatial distribution of peak particles, and (right) the distribution of non-peak particles. In the density profile panels, solid lines represent the density of peak particles, while dotted lines show the total density including both peak and non-peak components. Thin grey lines in the background display the density profiles of the same halo at all other snapshots, allowing comparison across time.

At the earliest time shown (top row, $a = 2.83$), both the peak and total density profiles follow a clean $r^{-3/2}$ power law, indicating the formation of the prompt cusp. The background grey curves stack tightly around the green profile, indicating a long-term stability prior to the merger. The density map of the peak particles (middle panel) shows a compact, centrally concentrated cusp, while the non-peak component (right panel) appears diffuse and loosely distributed around the halo. This reflects ongoing smooth accretion: non-peak material accumulates at large radii without disturbing the cusp, contributing only modestly to the outskirts of the total density profile, as seen in the slight offset between the dotted and solid green lines.

As the merger unfolds from the second row ($a = 3.31$) through the fifth row ($a = 5.10$), the halo undergoes progressive dynamical disruption. At $a = 3.31$, the total density profile (dotted green line) begins to rise at large radii, reflecting the influence of the infalling secondary halo whose mass is starting to contribute to the outer density structure. Meanwhile, the peak-only profile (solid green line) flattens and declines in the inner regions, likely due to a reflexive response of the cusp structure becoming elongated or displaced under the influence of the approaching halo’s gravitational field. 
At $a = 3.44$ and $a = 3.63$, the halo enters a highly non-equilibrium state due to the ongoing merger. Both the total and peak-only density profiles exhibit significant fluctuations as the secondary halo orbits the central system. This is reflected in the rightmost column, where the distribution of non-peak particles reveals distorted substructures associated with the infalling halo. The growing discrepancy between the total and peak-only density profiles in the central region indicates that the material from the secondary halo has sunk deep into the centre of the primary halo. At $a = 5.10$, the substructure previously visible in the non-peak particle distribution has been disrupted, and the system has stabilized into a more symmetric configuration, with the centre of the merged halo now fully overlapping that of the primary. 

The resulting system resembles a typical halo at redshift $z = 0$ (bottom panels). The total density profile begins to approach the NFW form, while the peak-only profile clearly deviates from the earlier prompt cusp configuration, reflecting the disruption of the initial self-similar structure. At $a = 8.00$, the halo at redshift $z = 0$ exhibits an NFW-like structure. The non-peak particle distribution has become symmetric, with no visible substructures, indicating that the system has reached a relaxed, equilibrium state. This outcome resembles the behaviour seen during mergers with artificial fragments (Section~\ref{sec:example_spurious}), both of which erase the central cusp structure.

\subsubsection{Disturbances from large-scale structures}
\label{sec:example_lls}
\begin{figure*}
    \centering
    \includegraphics[height = 0.9\textheight]{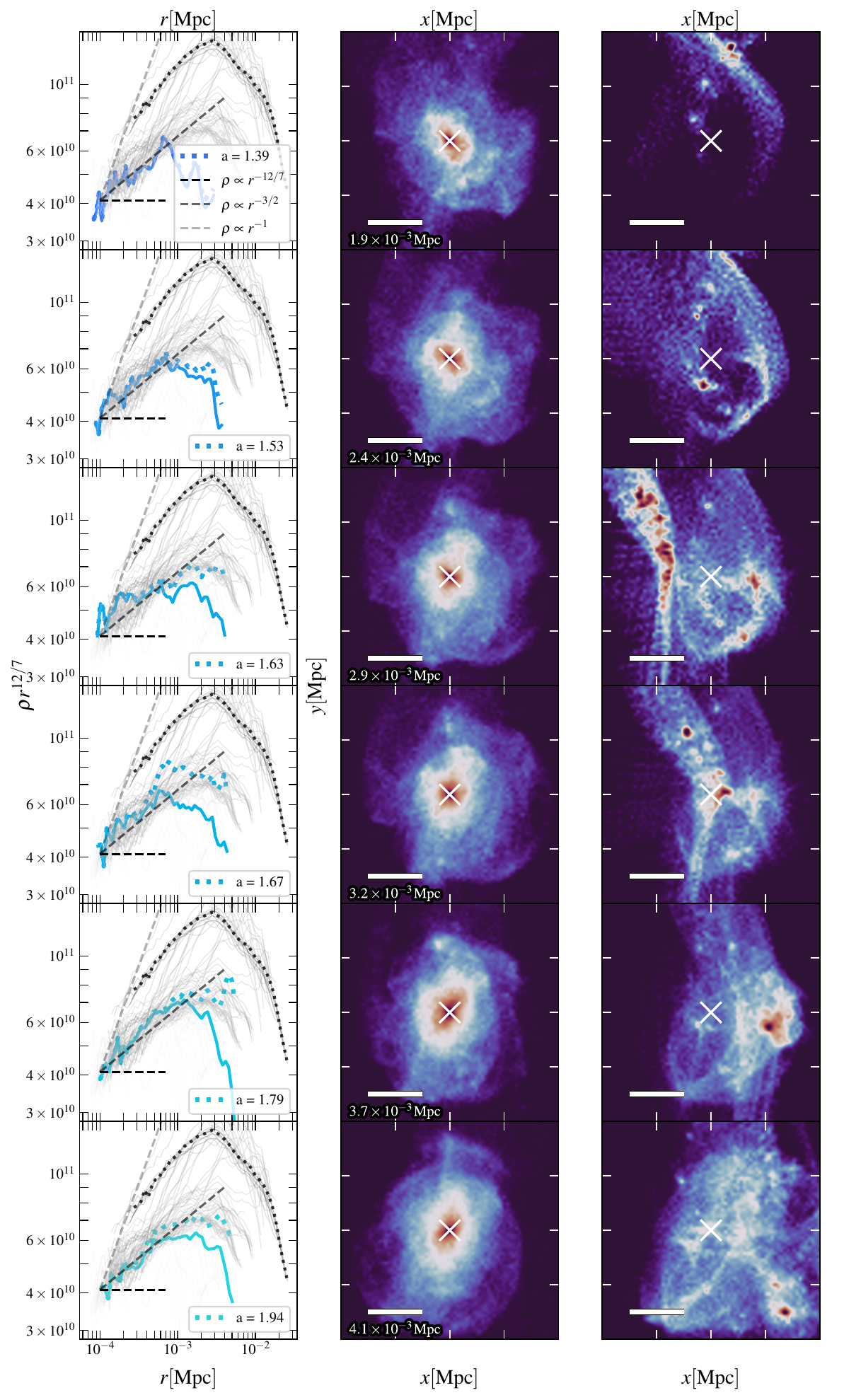}
    \caption{Evolution of the configuration and density profile versus physical radius $r$ of a halo interacting with a local large-scale structure. Snapshots are arranged in time from \textit{top} (before the interaction) to \textit{bottom} (after the interaction). Each row corresponds to a simulation snapshot, showing (\textit{left}) the spherically averaged density profile, (\textit{middle}) the spatial distribution of peak particles, and (\textit{right}) the distribution of non-peak particles. 
    The use of colours and symbols is the same as in Figure~\ref{fig:Evo_Merge}.
    This figure illustrates a transient encounter between a prompt cusp and a filamentary structure passing through the halo, which causes modest accretion and a temporary perturbation to the outer density profile, while preserving the inner cusp structure.}
    \label{fig:Evo_Filament}
\end{figure*}
Figure~\ref{fig:Evo_Filament} presents the evolution of a halo interacting with a nearby filamentary structure. The arrangement is the same as in Figure~\ref{fig:Evo_Merge}.

In the earliest snapshot ($a = 1.39$), the halo is in a quiescent stage. The total and peak density profiles are in close agreement across all radii. The rightmost panel reveals a coherent stream of non-peak particles around the outskirts of the halo, consistent with ongoing smooth accretion. As this infalling material has not yet penetrated the inner regions, the central cusp remains undisturbed. At $a = 1.53$, a coherent stream of non-peak particles becomes visible wrapping around the halo centre, as seen in the rightmost panel. Despite this ongoing accretion, the central cusp remains largely undisturbed, with the peak-only density profile preserving its original shape. A slight increase in the total density profile, particularly at larger radii, reflects the contribution of smoothly accreted material. Between $a = 1.63$ and $a = 1.67$, a filamentary structure approaches and intersects the halo, as revealed in the non-peak particle maps. The filament passes directly through the halo centre, bringing with it several dense clumps. In response, the total density profile exhibits a noticeable rise at large radii. However, the inner slope continues to align closely with the peak-only profile which retains its power-law form, indicating that the central structure remains largely undisturbed by the interaction, with the halo centre still dominated by peak particles. By $a = 1.79$, the filament begins to detach from the halo, as seen in the rightmost panel. Much of the infalling material either disperses or escapes the halo. The density profile at large radii falls accordingly, indicating that the newly accreted material is no longer contributing to the halo potential. Despite the transient disturbance, the inner profile remains stable, and the peak particles continue to dominate the central structure. At the final snapshot ($a = 1.94$), the interaction has fully subsided. The non-peak material is now more diffusely distributed, and no clear substructures remain. The total and peak density profiles have largely re-converged, especially at small radii, indicating that the halo has returned to a relatively relaxed state. Notably, a discrepancy between the profiles begins to emerge as the peak density profile flattens at larger radii, marking the boundary of the prompt cusp and the outer limit of the region dominated by peak particles. In the end, the cusp remains intact, consistent with the relatively weak and transient nature of the disturbance. This outcome resembles the behaviour seen during minor mergers (Figure~\ref{fig:Evo_Profile}) and contrasts sharply with the significant disruption caused by major merger (Sec.~\ref{sec:example_merger}) and mergers with artificial fragments (Sec.~\ref{sec:example_spurious}).

\subsubsection{Mergers with artificial fragments}
\label{sec:example_spurious}
\begin{figure*}
    \centering
    \includegraphics[width = \textwidth]{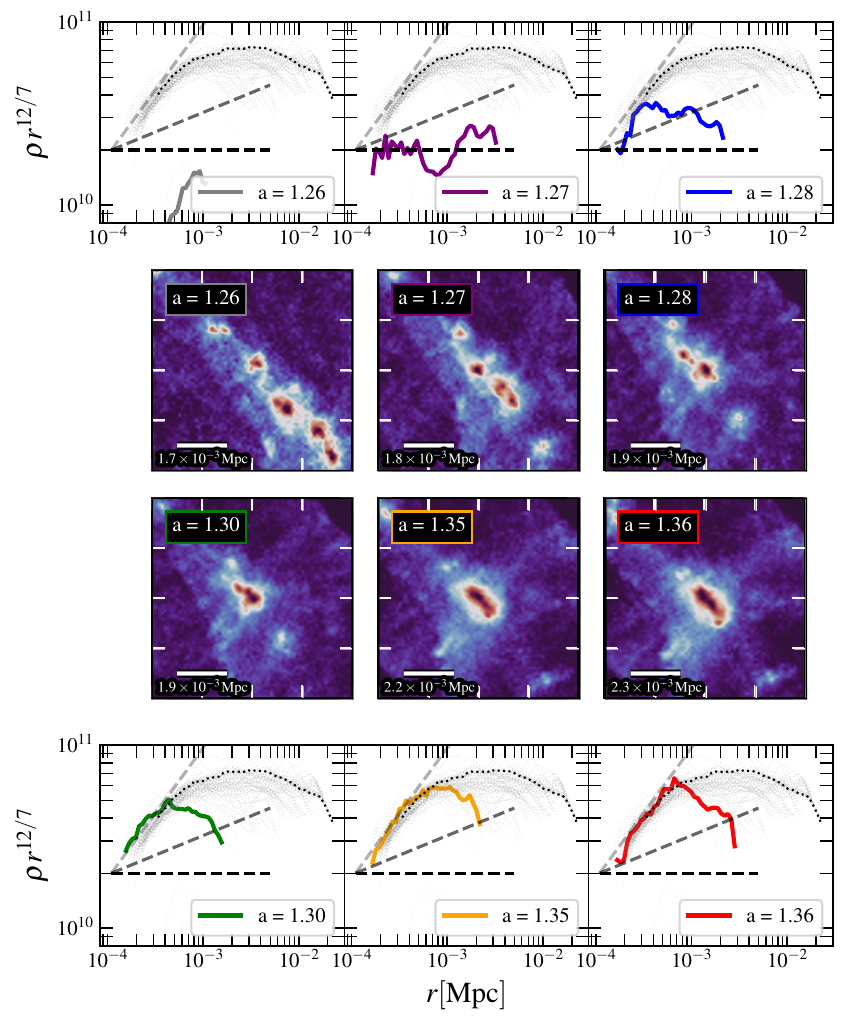}
    \caption{Evolution of the configuration and density profile versus physical radius $r$ of a halo undergoing a merger with an artificial fragments. Snapshots are arranged in time from left to right, and from top to bottom. The corresponding density profiles for all halo particles and peak particles are plotted in the middle in dotted and solid lines, respectively. The thin grey lines represent the full sequence of density profiles throughout the simulation. The $y$ axis of the density profile is rescaled by $r^{12/7}$ as Figure~\ref{fig:Evo_Profile}.}
    \label{fig:Evo_Spurious}
\end{figure*}
Artificial fragments are prevalent in our simulations, particularly in haloes near the cutoff of the power spectrum (see Section~\ref{subsec:spurious}). These artificial structures can form early and close to the proto-halo, even before the prompt cusp has stabilized. Figure~\ref{fig:Evo_Spurious} illustrates one such case, in which a forming prompt cusp is disrupted by early-formed artificial substructures.
Throughout the panels, the peak and total density profiles (solid and dotted lines, respectively) are indistinguishable. This is because the artificial substructure forms from material that originally belonged to the primordial peak itself. As such, the particles involved are classified as peak particles by our algorithm and are not excluded.

In the early snapshot at $a = 1.26$, three compact, high-density clumps - identified as artificial fragments - are already visible in the density map. The central density profile is sharply truncated, showing little to no buildup of the expected prompt cusp structure. By $a = 1.27$ and $a = 1.28$, these artificial fragmentats begin interacting and merging with one another, gradually converging near the centre of the forming halo. The increasing density profiles (top row) exhibit noisy, irregular features due to the ongoing merger. By $a = 1.30$, the clumps have largely coalesced into a single central object. The density profile begins to resemble that of the final halo at redshift $z = 0$. At $a = 1.35$ and $a = 1.36$, continued accretion and the residual orbital motion of the merging clumps result in an elongated and dynamically unrelaxed central structure. Despite this, the overall density profile remains consistent with the NFW form. Comparison with the profile at redshift $z = 0$ suggests that subsequent evolution does not significantly alter the inner structure but instead gradually drives it toward the canonical NFW shape - supporting the interpretation of the NFW profile as a dynamical attractor for virialized haloes.

The resulting central disturbance is qualitatively similar to the disruption seen in major mergers (cf. Figure~\ref{fig:Evo_Profile}). This suggests that mergers with artificial clumps can deliver sufficient mass and central energy to fully disrupt the prompt cusp, despite the spurious origin of the perturbation.
Ultimately, this example reinforces the idea that significant clumpy accretion events - regardless of whether it originates from real or artificial structures - can significantly alter or even destroy the prompt cusp if it occurs before full collapse.

\subsection{Density profile evolution}
\label{subsec:fitting}
\begin{figure*}
    \hspace*{-1cm}
    \centering
    \includegraphics[width = 1.1\textwidth]{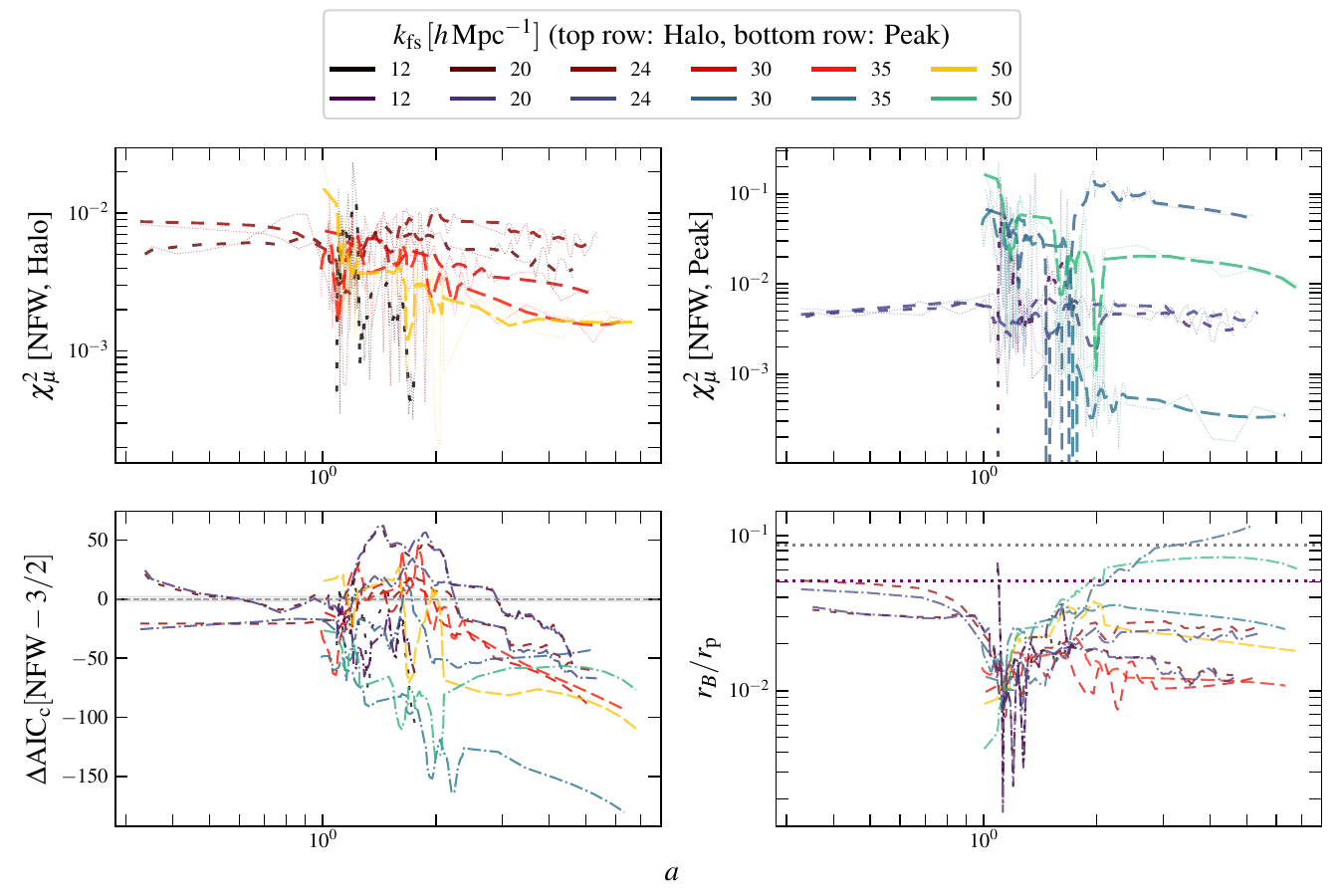}
    \caption{Evolution of the statistical fitting result of halo density profile for a representative halo (index 206). \textit{Top row}: smoothed $\chi^2_\mu$ to an NFW profile for the full halo (\textit{left}) and for peak-selected material (\textit{right}) versus rescaled time $a_{\mathrm{c}}$, the collapse time of the matched primordial peak. The shaded intervals mark epochs when an NFW fitting is worse than the $3/2$ model, including the epochs when NFW fitting fails. \textit{Bottom-left}: relative preference between the NFW and $3/2$ model. \textit{Bottom-right}: evolution of the characteristic inner scale $r_B/r_{\mathrm{p}}$ tracing how far the inner power law extends, rescaled by the characteristic size of the primordial peak $r_{\mathrm{p}}$. Colours encode $k_{\rm fs}$ as indicated in the title panel; solid curves correspond to halo profile and dash-dot curves to peak profile. }
    \label{fig:example_fitting}
\end{figure*}
The diagnostics presented in this section show how the inner density structure of our haloes evolves from an early prompt–cusp phase towards an NFW-like configuration. We quantified (i) how well the profiles are described by an NFW fit, (ii) when an NFW form is statistically preferred over fixed power-law models, (iii) the comparison between the performance of the empirical $3/2$ profile and the self-similar prediction $12/7$ profile, and (iv) whether a more flexible self–similar prescription that accounts for the Lagrangian peak shape, $\lambda(\gamma)$, provides any improvement. 
We begin by quantifying the density profile evolution with the fitting statistics $\chi_\mu2$ and $\Delta\mathrm{AIC}_\mathrm{c}$ calculated in Appendix~\ref{app:profile_fitting}. Complete per-halo diagnostic, are provided in Appendix~\ref{app:result}, here we summarise the key definitions and present representative results.

\subsubsection{The time evolution of NFW fitting}

In the upper panels of Figure~\ref{fig:example_fitting} we show the evolution of the reduced
chi-squared, $\chi_\mu^2$, of NFW fits to both the full halo and the peak-only density profiles, plotted against the rescaled time $a$.
The result of all halos are shown in Fig~\ref{fig:NFWchi2}.
For each component, curves are coloured by the free-streaming wavenumber $k_{\rm fs}$, with a red-to-yellow gradient for haloes and a purple-to-green gradient for peak-only profiles, while different line styles distinguish individual $k_{\rm fs}$ values. The raw $\chi_\mu^2$ measurements are smoothed using a Savitzky-Golay filter to highlight the underlying trends, with the unsmoothed values shown as faint dotted lines in the background. 

At later times, most haloes display a gradual and smoother decrease of $\chi_\mu^2$, approaching a quasi-stationary regime in which the NFW fit quality stabilises. This trend indicates that the inner density profiles of the haloes become progressively more consistent with the NFW form as they evolve away from collapse. 

Just after collapse, both the halo and peak components exhibit a relatively noisy and non-monotonic evolution in $\chi_\mu^2$, with especially strong excursions in the peak-only profiles. This suggests that the inner structure is strongly perturbed during this early phase, for example by mergers with artificial fragments that inject clumpy material into the prompt cusp. At fixed time, the peak-only density profiles almost always yield larger $\chi_\mu^2$ than the corresponding full halo profiles. This demonstrates that subtracting the material accreted after the collapse of the primordial peak breaks the apparent universality of the NFW profile - this suggests that the NFW profile is a result of haloes evolving with environmental interaction.

The $\chi^2_\mu$ result exhibits sharp drops in $\chi_\mu^2$ at intermediate times, visible simultaneously in both the halo and peak-only components. These rapid improvements in the NFW fit suggest episodes of violent relaxation associated with major mergers, which efficiently redistribute material and drive the system towards an NFW-like inner profile. The timing of these events is similar across all $k_{\rm fs}$ realisations for a given halo, since the large-scale structure is preserved and major mergers are not fully suppressed. However, the responses of the halo and peak profiles to such events are diverse.

\subsubsection{The evolution from prompt cusp to NFW}
While the evolution of $\chi_\mu^2$ in the upper panels in Figure~\ref{fig:example_fitting} already shows that haloes tend to become more NFW-like over time, it does not by itself indicate at which stages a prompt-cusp–like power law provides a better description than NFW. To address this, we use the AICc-based model comparison introduced in Sec.~\ref{subsec:self-similar} to track, for each halo and free-streaming scale, whether the inner profile is more consistent with an $r^{-3/2}$ cusp or with an NFW profile.
The bottom left panel in Figure~\ref{fig:example_fitting} presents the evolution of the AICc difference, $\mathrm{AIC}\mathrm{c}[{\rm NFW}] - \mathrm{AIC}\mathrm{c}[{3/2}]$, as a function of $a$ for both the full halo and the peak-only density profiles. The difference being smaller than 0 means that NFW is a better fit, and being larger than 0 means that the power-law profile is a better fit. 

Just as in the NFW $\chi^2$ results, all profiles show a general decline towards negative values of $\Delta\mathrm{AIC}\mathrm{c}$ with time - especially the fact that the earliest snapshots do not have a valid NFW fit (grey shaded regions in Figure~\ref{fig:NFW_32}) is accounted - indicating that the haloes are better described by a power-law density profile at early times and become increasingly well described by NFW at later times. At the moment an NFW fit first becomes available, it is already preferred over the $-3/2$ model in most cases. The main difference between the halo and peak density profiles persists at late times, when the systems are accreting predominantly smoothly.

We identify the prompt-cusp phase by selecting all snapshots in which the $r^{-3/2}$ model outperforms NFW (i.e. $\mathrm{AIC}_\mathrm{c}[\rm NFW]>\mathrm{AIC}_\mathrm{c}[{3/2}]$) including epochs where the NFW fit fails. Our results show that the transition to an NFW–like profile generally occurs before $a \simeq 3$. 

\subsubsection{Size of prompt cusps}
With the bottom right panel of Figure~\ref{fig:example_fitting} we discuss the characteristic scale of the prompt cusps. With \(B\) denoting the number of inner logarithmic bins of the best fitting power law model (Sec.~\ref{subsec:fitting}); the corresponding radius \(r_B\) naturally marks the largest radius within which the density profile is well described by a the power law profile and can therefore be interpreted as the characteristic size of the prompt cusp. 
We plot \(r_B\) for both the halo and peak profiles using the \(r^{-3/2}\) fit, and rescale by the size of the matched primordial peak, \(r_{\rm p}\) (Sec.~\ref{subsec:self-similar}). According to \citet{White2022}, the theoretical second apocentre radius, scaled by \(r_{\rm p}\), is a function of the enclosed mass fraction $f_{\rm p}$ of the total mass in the peak. For reference we overlay the values from their Eq.~(14) for \(f_{\rm p}=1\) and \(f_{\rm p}=0.5\) as horizontal dashed lines in grey and purple, respectively.

The characteristic size of the prompt cusp increases after peak collapse and grows as the prompt cusp forms. However, most haloes do not attain the full mass prompt cusp regime ($f_{\rm m}\simeq 1$). This is because of two reasons: the outer edge of the prompt cusp departs from a pure power law as additional material accumulates and cusp is disrupted before fully forming by mergers with artificial fragments, causing the \(r^{-3/2}\) preference over NFW to end prematurely. The results of all halos are in Figure~\ref{fig:32_Size}.

\begin{figure}
    \centering
    \includegraphics[width=\columnwidth]{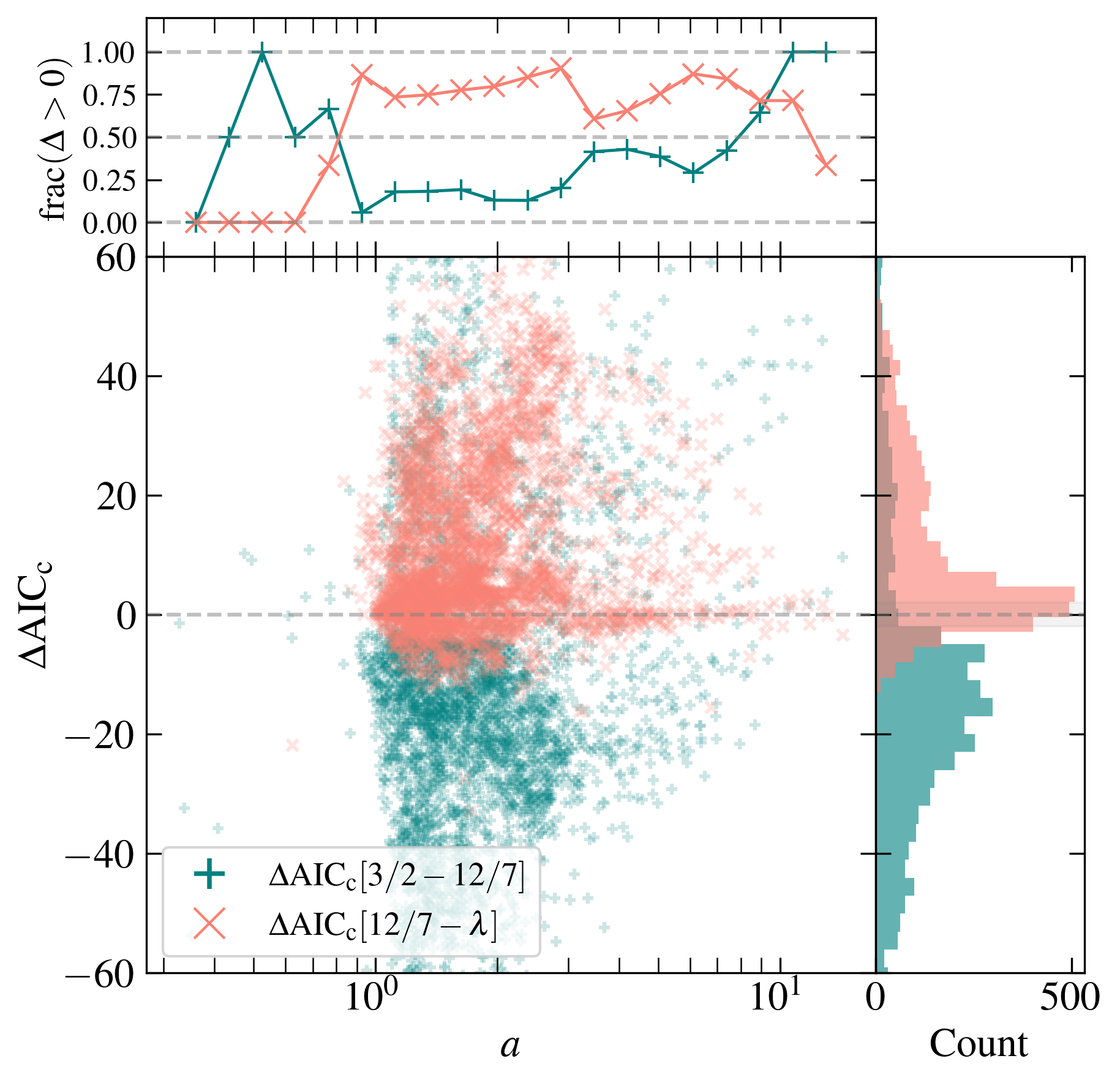}
    \caption{Model comparison of the \textit{peak} density profile across all haloes, snapshots, and $k_{\rm fs}$, restricted to epochs where $\Delta\mathrm{AIC}_\mathrm{c}[{\rm NFW} -{3/2}]>0$ (i.e. the $-3/2$ power law outperforms NFW), together with epochs where the NFW fit fails. Main panel shows $\Delta\mathrm{AIC}_\mathrm{c}[3/2-12/7]$ as $+$ and $\Delta\mathrm{AIC}_\mathrm{c}[12/7-\lambda]$ as $\times$ with \textit{x} axis as $a$ where $a$ is the rescaled scale factor. For both of the diagnostics, positive values of $\Delta\mathrm{AIC}_\mathrm{c}[M_1 - M_2]$ favour the model $M_1$, while negative values favour the model $M_2$. The top left panels shows the fraction of $\Delta\mathrm{AIC}_\mathrm{c}[M_1 - M_2] > 0$, with same colour and sign as the main panel. The side panel shows the distributions of the two $\Delta\mathrm{AIC}_\mathrm{c}$ samples (horizontal histograms) on the same $y$-axis scale as the main panel. 
    }
    \label{fig:theory_compare_peak}
\end{figure}

\subsubsection{Theoretical models of the halo density profile}
\label{subsubsec:models}
 Previous works based on simulation results found a systematically shallower density profile of the early haloes \citep{2005Diemand,2010Ishiyama,2014Ishiyama,2013Anderhalden,2019Delos,2023Delos,Ondaro2024}, with a characteristic value of $\lambda \simeq 3/2$ compared to the theoretical prediction of $\lambda = 12/7$ from self-similar collapse models \citep{White2022}.

We now compare directly the three power-law models using the $\Delta{\rm AICc}$ statistic introduced in App.~\ref{app:profile_fitting}. 
Figures~\ref{fig:theory_compare_peak} shows, for peak profiles, $\Delta\mathrm{AIC}_\mathrm{c}[3/2{-}12/7]\equiv \mathrm{AIC}_\mathrm{c}[{3/2}]-\mathrm{AIC}_\mathrm{c}[{12/7}]$ as teal “$+$” symbols and $\Delta\mathrm{AIC}_\mathrm{c}[12/7{-}\lambda]\equiv \mathrm{AIC}_\mathrm{c}[{12/7}]-\mathrm{AIC}_\mathrm{c}[{\lambda}]$ as salmon “$\times$”, plotted against the rescaled time $a$. The right–hand panels display the corresponding $\Delta\mathrm{AIC}_\mathrm{c}$ distributions, while the top panels report, in the $a$ bins, the fraction of snapshots with $\Delta\mathrm{AIC}_\mathrm{c}>1$ favouring the first model in each pair (i.e. $3/2$ over $12/7$, and $12/7$ over $\lambda$). For ease of comparison, we draw three dashed grey horizontal reference lines at 0, 0.5 (two profiles being equally favoured) and 1 in the upper panel.


In Figure\ref{fig:theory_compare_peak}, the $\Delta\mathrm{AIC}_\mathrm{c}[3/2{-}12/7]$ distributions are clearly skewed toward values $<0$ in both figures, indicating a systematic preference for the $3/2$ cusp over the theoretical self–similar $12/7$ slope during the prompt–cusp phase, consistent with prior simulation results. At later times, as the central slope shallows toward NFW and the halo departs from any single power law, the advantage of $3/2$ diminishes and $12/7$ profile is favoured again due to it being closer to the NFW inner density slope -1. 

We find that the majority of $\Delta\mathrm{AIC}_\mathrm{c}[12/7-\lambda]$ points lie close to 0, with an enhanced distribution toward positive values, implying that the peak-informed variable-slope $\lambda$ profile improves upon the fixed 12/7 profile. 

The patterns are clearer in the peak–profile fits (the halo results are shown in Figure~\ref{fig:theory_compare_halo}), consistent with our interpretation that prompt cusps in the full halo are rapidly erased after collapse, while peak–selected profiles evolve more gradually and often retain a cusp–like form. The model ranking flips at roughly the same rescaled time indicated by $\Delta\mathrm{AIC}_\mathrm{c}[3/2{-}12/7]$: during the prompt–cusp window the $3/2$ profile outperforms $12/7$ and the variable–$\lambda(\gamma)$ model likewise outperforms $12/7$; beyond this window, both preferences reverse.

\section{Discussion}

Although it is widely accepted that the NFW profile emerges from the hierarchical assembly of dark matter, its precise physical origin remains a subject of debate. 
The evolution is typically described in terms of either accretion or merging. Accretion refers to the overall \textit{quantity} of mass added to a halo over time, while merging describes the \textit{form} of that infall takes, involving discrete encounters with collapsed structures.
These two concepts capture distinct but correlated aspects of halo growth: encounters with massive objects naturally contribute to both the mass accretion rate increase and the merger history. To distinguish between diffuse and clumpy infall, the term smooth accretion is often used to describe the gradual incorporation of diffuse material. 
The relative importance of mergers versus smooth accretion in determining the internal structure of haloes and in particular, in producing the NFW profile - has been the focus of ongoing theoretical and simulation-based studies:

Two approaches have been developed to explain the emergence of the NFW profile from hierarchical structure formation: analytical models based on first principle physical mechanisms, and empirical models grounded in simulation data.
Analytical models aim to derive the NFW profile from the initial condition within a hierarchical framework. Early work combined the secondary infall model with statistical descriptions of smooth accretion from the neighbourhood Gaussian random peaks \citep{1986Bardeen} onto the local maximum \citep{1985Hoffman}. These models were later extended to include additional physical processes such as tidal stripping, dynamical friction, and the heating of parent haloes by infalling substructures \citep{1999Nusser}, as well as angular momentum generation by secondary perturbations \citep{1987Ryden, 2004Williams}. In parallel, \citet{1998Syer} demonstrated that repeated mergers, combined with dynamical friction and tidal stripping, adds the material from the satellites to the parent halo, driving haloes toward an universal density profile with an NFW-like central cusp, suggesting that the NFW form may act as a dynamical attractor.

Empirical approaches assume the NFW form and focus on reproducing the statistical distribution of its parameters, particularly the concentration $c = r_{200}/r_s$.
Several studies have proposed a two-phase picture of halo formation, in which an early, rapid accretion phase - often dominated by major mergers - sets the inner density profile through violent relaxation, followed by a slower phase of smooth accretion that builds up the outskirts \citep{2002Wechsler, 2003Zhao, 2009Zhao}. In this framework, the NFW profile arises naturally from the time-integrated effects of these two phases. More recently, \citet{2013Ludlow, 2014Ludlow, 2016Ludlow} demonstrated that when halo density profiles are rescaled using the Mass Accretion History (MAH), the NFW profile is recovered, suggesting that the profile is closely tied to the halo’s growth history.

Taking the self-similar collapse models of \citet{Fillmore1984}, \citet{1985Bertschinger}, and \citet{White2022} as representations of early halo formation, one can interpret the observed universality at both the initial and final stages of halo growth as implying some degree of universality in the growth process itself. This, in turn, suggests that the interaction between the primordial peak and its environment - both direct accretion and environmental influences - could follow broadly similar patterns across haloes. It is natural to assume that the same physical processes responsible for destroying prompt cusps may also play a key role in shaping the final NFW structure.

The apparent transformation from a prompt cusp to an NFW-like centre proceeds through two qualitatively different channels. Smooth inflow adds centrally concentrated material without strongly mixing the innermost region; the composite density profile then fits an NFW-like form because the added envelope overshadows the original cusp \citep[cf.][]{2019Delos,Delos2023}, even though the peak-only component still retains an $r^{-3/2}$-type profile. By contrast, mergers can drive rapid potential fluctuations and replace the original prompt cusp with a new central structure whose inner slope is closer to $-1$ \citep[cf.][]{2017Angulo,2018Ogiya}. Early numerical fragments can mimic this second pathway if not removed, producing an apparent early transition \citep{Ondaro2024}. The survival of prompt cusps therefore depends primarily on the mode and timing of mass addition and on the strength and duration of environmental interactions.

Across our sample, we find that haloes undergoing very different evolutionary pathways - including smooth accretion, mergers with genuine subhaloes, and disruption by artificial fragments - all converge to NFW-like density profiles at late times. In some cases the central prompt cusp is preserved and merely buried beneath an accreted envelope, whereas in others it is strongly perturbed or even destroyed, we refer to this as burial rather than destruction. Nevertheless, the final profiles are consistently well described by the NFW form. 
The simulations in which the halo structure is strongly affected by early artificial fragmentation also end up NFW-like at $z = 0$ after undergoing several major mergers. This indicates that, once an NFW-like central profile has been established, it is remarkably stable to subsequent major mergers, in line with earlier idealised studies that argued for the dynamical robustness of NFW profiles. This supports the view that NFW behaves as an attractor, reached through multiple physical mechanisms that partially erase the detailed imprint of the primordial peak. We emphasise that, although mergers with genuine haloes and interactions with artificial fragments can have qualitatively similar effects on the evolution of the spherically averaged density profile, this does not imply that their internal structures are alike. In particular, the artificial fragments lack the physically self-consistent phase-space structure of genuine haloes, so the way they perturb the gravitational potential within the prompt cusp is fundamentally different, even if the eventual macroscopic outcome - an NFW-like profile - appears similar.

In the smooth accretion dominated cases, where the prompt cusp survives at the centre, our results suggest that it may be possible in principle to describe the infalling material with an universal profile superposed on the primordial cusp. However, such a description would not apply in cases where the central cusp is disrupted. 

As noted by \citet{White2022}, the offset between the analytic prediction $\gamma=-12/7$ and the slopes measured in simulations can arise from the anisotropic, triaxial nature of collapse: the three principal axes collapse at different times, which invalidates a single, uniform self-similar rescaling. 
We partially addressed this by incorporating information on the Lagrangian peak shape $\gamma$ beyond the isotropic second-order expansion, enabling a variable slope fit $\lambda(\gamma)$ using the ellipsoidal radius $r_{\rm ell}$. Our model comparison demonstrated that this approach yielded a modest but systematic improvement over the fixed $\lambda = -12/7$ prediction. Nevertheless, the persistent statistical preference for the shallower $\rho \propto r^{-3/2}$ cusp in early phases indicates that the issue of triaxial collapse and non-self-similar evolution is not fully resolved by our current $\gamma$-based approach.
A comprehensive assessment will require an explicit treatment of the cusp's triaxial geometry; we defer a detailed shape analysis to future works.

\section{Conclusion}

In this study, we investigated the formation and subsequent evolution of prompt cusps in dark matter haloes in cosmological environments, using a suite of 64 zoom-in $N$-body simulations. 
By varying the free-streaming cut-off, we controlled the abundance of small-scale structure while keeping the large-scale environment fixed.

Our results confirm that prompt cusps - power-law inner density profiles with index of 1.5 - emerge universally following the collapse of primordial density peaks. These cusps form rapidly and are well-described by self-similar solutions, consistent with previous simulations \citep[e.g.][]{Delos2023}. 

We described a new algorithm for identifying artificial fragments - spurious substructures seeded by discreteness noise - using a physically motivated, peak-based diagnostic that can be applied consistently across different free–streaming scales without retuning for each $k_{\rm fs}$ at fixed resolution. For each subhalo we traced its Lagrangian particle set back to the initial conditions and quantified its association with the catalogue of linear density peaks by means of a structural overlap statistic, classifying subhaloes that match at least one peak above the collapse threshold as genuine and the remainder as artificial. We then used this classification to calibrate simple, two-parameter cuts in the plane of subhalo maximum progenitor mass and Lagrangian axis ratio, extending the empirical approach of previous WDM studies. By scanning this parameter space we computed the true positive rate (TPR), false positive rate (FPR), and Youden’s $J$ statistic for each choice of cut, and for every $k_{\rm fs}$ we reported three operating points tailored to different scientific priorities: a `Max-$J$' solution that maximises the overall discrimination between genuine and spurious objects, a `High-TPR' solution that favours completeness of genuine subhaloes at the cost of a higher contamination fraction, and a `Low-FPR' solution that yields a particularly clean sample of genuine haloes with minimal inclusion of artificial fragments. As the optimal parameter values are consistent across our simulations with different $k_{\rm fs}$, these calibrated choices should be applicable to other WDM-like simulations with comparable mass resolution. If resolution differences are properly accounted for through $M_{\rm lim}$ (Eq.~\ref{eq:M_lim}), then, in principle, the same cuts can be extended to simulations with different resolutions and mass scales.

We introduced an improved self-similar prescription that accounts explicitly for the shape of the Lagrangian peak. By fitting the linear density field around each primordial maximum with an ellipsoidal quadratic form, we measured a shape parameter $\gamma$ and used it to construct a variable-index self-similar solution with slope $\lambda(\gamma)$ and corresponding amplitude $A(\gamma)$. The AICc comparison between this model and the fixed $12/7$ solution shows that allowing $\lambda$ to respond to the measured peak shape yields a modest but systematic improvement both during and after the prompt cusp phase, particularly for the peak profiles. In other words, part of the tension between the self-similar prediction and the $r^{-3/2}$ cusps seen in simulations can be traced back to the fact that realistic Lagrangian peaks deviate from the idealised, spherically symmetric second-order expansion assumed in previous analytic work. The remaining offset must then be attributed to genuine departures from self-similarity assumption itself - broken by smooth accretion, mergers and tidal perturbations.

On a population level, our statistical fitting of the density profile provides a coherent picture of how the inner density structure of our haloes evolves from an early prompt-cusp phase towards an NFW-like state. First, by tracking the reduced $\chi^2_\mu$ of NFW fits, we showed that the inner profiles of the halo \emph{as a whole} becomes progressively better described by NFW with time, whereas the profile of the particles only associated with the initial peak systematically retains larger $\chi^2_\mu$ values at fixed epoch. This demonstrates that the apparent universality of the NFW form is largely a property of the total mass distribution - including later accretion - rather than of the primordial peak material alone. Secondly, the AICc comparison between NFW and fixed power-law models reveals that, immediately after collapse and before a stable NFW fit can be obtained, the inner profiles are better described by a single power law with slope close to $-3/2$. As soon as an NFW fit becomes statistically meaningful, however, the NFW model is almost always preferred over the $r^{-3/2}$ profile, signalling a rapid transition from a prompt cusp like configuration to an NFW inner structure. Third, by directly contrasting the $r^{-3/2}$ and $r^{-12/7}$ models, we confirm that the prompt cusps measured in our simulations are systematically shallower than the analytic self-similar prediction with slope $-12/7$, although adopting a more accurate description of the primordial peak shape yields a modest improvement in the fits.

We demonstrate that our peak-particle selection is physically meaningful by examining representative examples and reviewing their spatial distribution in detail. From the group-level picture, we identify three distinct evolutionary pathways by which prompt cusps transform into NFW-like haloes: (1) Major mergers with comparable-mass haloes produce strong dynamical disturbances that disrupt the self-similar structure and redistribute central material. Both the peak density profile and the halo density profile depart from the power-law profile, and the halo density profile rapidly evolves towards NFW profile. (2) Interactions with large-scale filaments, which brings minor mergers and an increased smooth mass accretion, resulting in modest perturbations in the halo outskirts. In such cases, the inner prompt cusp remains largely intact and continues to dominate the central gravitational potential. The resulting peak density profile preserves its power-law configuration, while being embedded inside a halo density profile that deviates from the power-law. (3) Mergers with artificial fragments have an influence on the density profile similar to that of a major merger. When these artificial structures merge with the central halo prior to full cusp formation, they induce an effect on the matter distribution similar to the major merger and result in an early transition to an NFW profile. 

The mode and timing of mass accretion therefore govern cusp survival. Violent, early, and clumpy events tend to disrupt prompt cusps, while smooth accretion tends to preserve them. The observed convergence of disrupted haloes toward NFW-like profiles further supports the interpretation of the NFW form as a dynamical attractor in hierarchical structure formation. 

Limitations and future work include extending the analysis across a broader range of halo masses and cosmologies, increasing simulation resolution to suppress artificial fragmentation, and quantifying the response of halo structure to environmental perturbations.

\section{Acknowledgements}
We thank Sten M. Delos for meaningful discussions and Simon White for useful comments on the importance of artificial fragments. YW is supported by the Durham University Doctoral Scholarship and the Discretionary Stipend. SB is supported by the UKRI Future Leaders Fellowship (grant numbers MR/V023381/1 and UKRI2044). ARJ is supported by the STFC consolidated grant ST/X001075/1. This work used the DiRAC@Durham facility managed by the Institute for Computational Cosmology on behalf of the STFC DiRAC HPC Facility (www.dirac.ac.uk). The equipment was funded by BEIS capital funding via STFC capital grants ST/K00042X/1, ST/P002293/1, ST/R002371/1 and ST/S002502/1, Durham University and STFC operations grant ST/R000832/1. DiRAC is part of the National e-Infrastructure.

\section{Data Availability}
The data presented in this paper will be shared upon reasonable
request to the corresponding author.







\bibliographystyle{mnras}
\bibliography{reference}


\appendix
\section{Density profile fitting}
\label{app:profile_fitting}

\subsection{Fitting method}
The fitting is done in logarithmic form, where the NFW model can be written as
\begin{equation}
    \log_{10}\rho(r)
    = \log_{10}\rho_{\mathrm{s}}
      - \log_{10}(r/r_{\mathrm{s}})
      - 2\,\log_{10}\!\left(1 + r/r_{\mathrm{s}}\right),
\end{equation}
where $\rho_{\mathrm{s}}$ and $r_{\mathrm{s}}$ are free parameters.

All fits are performed using the innermost $B$ radial bins starting from the convergence radius $r_{\rm conv}$ (Sec.~\ref{subsec:self-similar}), with $B$ scanned from 20 up to the number of bins within the caustic radius $r_{\rm c}$. This choice ensures that the rapid slope variation near the caustic radius does not bias the fit, and that the best-fitting radius corresponds to the range where the prompt cusp is preserved. As a goodness-of-fit measure we compute the reduced $\chi^2_\mu$:
\begin{equation}
    \chi^2_\mu \;=\; \frac{\mathrm{SSE}}{B-F},
\end{equation}
where $F$ is the number of free parameters in the model and:
\begin{equation}
    \mathrm{SSE}
    = \sum_{i=1}^{B}
      \left[\log_{10}\rho_i
            - \log_{10}\rho_{\rm model}(r_i)
      \right]^2
\end{equation}
is the sum of squared residuals between the measured densities $\rho_i$ and the model prediction $\rho_{\rm model}(r_i)$.

We use the corrected Akaike Information Criterion, $\mathrm{AIC}_\mathrm{c}$, to compare models with different functional forms. For a fit to $B$ radial bins with sum of squared residuals $\mathrm{SSE} = \sum_{i=1}^m \left[\log_{10}\rho_i - \log_{10}\rho_{\rm model}(r_i)\right]^2$
and $F$ free parameters, we define:
\begin{equation}
  \mathrm{AIC}_\mathrm{c}
  = B \log_{10}\!\left(\frac{\mathrm{SSE}}{B}\right)
    + 2F
    + \frac{2F(F+1)}{B - F - 1}.
\end{equation}
Lower values of $\mathrm{AIC}_\mathrm{c}$ indicate a better balance between goodness of fit and model complexity. We use $\mathrm{AIC}_\mathrm{c}[\mathrm{NFW}]$ for the NFW fit, $\mathrm{AIC}_\mathrm{c}[3/2]$ for the fixed power-law model with inner slope $\lambda=3/2$, $\mathrm{AIC}_\mathrm{c}[12/7]$ for the fixed self-similar slope $\lambda=12/7$, and $\mathrm{AIC}_\mathrm{c}[\lambda]$ for the variable self-similar model whose slope depends on the Lagrangian peak shape.
In addition, we recompute these quantities to the peak profiles defined in Sec.~\ref{subsec:self-similar}.

For a pair of models $M_1$ and $M_2$ we consider the AICc difference:
\begin{equation}
  \Delta\mathrm{AIC}_\mathrm{c}[M_1 - M_2]
  \equiv \mathrm{AIC}_\mathrm{c}[M_1] - \mathrm{AIC}_\mathrm{c}[M_2].
\end{equation}
A positive $\Delta\mathrm{AIC}_\mathrm{c}$ implies that $M_2$ is preferred over
$M_1$, while a negative value favours $M_1$. For clarity, we denote the AICc values of different models using square brackets. 

\subsection{Fitting results}
\label{app:result}
This section provides per-halo overview across all simulations, assembling the key diagnostic plots that support the main results in Sec.~\ref{subsec:density_profile}.

In Fig~\ref{fig:NFWchi2} we show the evolution of the reduced chi-squared of NFW fits to both the full halo and the peak-only density pro-
files for all haloes. The figure is arranged as a $4\times 4$ grid in which each row corresponds to a pair of haloes: the left and right panels display the full halo profile and the peak-only profile respectively.
In every panel we overplot the complementary component as a faint colored background curve, allowing a direct visual comparison between the fit quality for the total halo and for the material associated with the primordial peak alone. 
Figure~\ref{fig:NFWchi2} shows that, in all haloes and for all free-streaming scales, the reduced chi-squared of the NFW fits generally decreases with time. 
Together, these trends support a picture in which early, clumpy accretion of the artificial fragments and mergers at late times perturb and sometimes destroy the prompt cusp, while subsequent smooth accretion drives the system towards an NFW attractor and highlights. Meanwhile, the peak-only density profile remains more stable against the smooth accretion, as seen for haloes~305, 337, and 443.
\begin{figure*}
    \hspace*{-1cm}
    \centering
    \includegraphics[width = 1.1\textwidth]{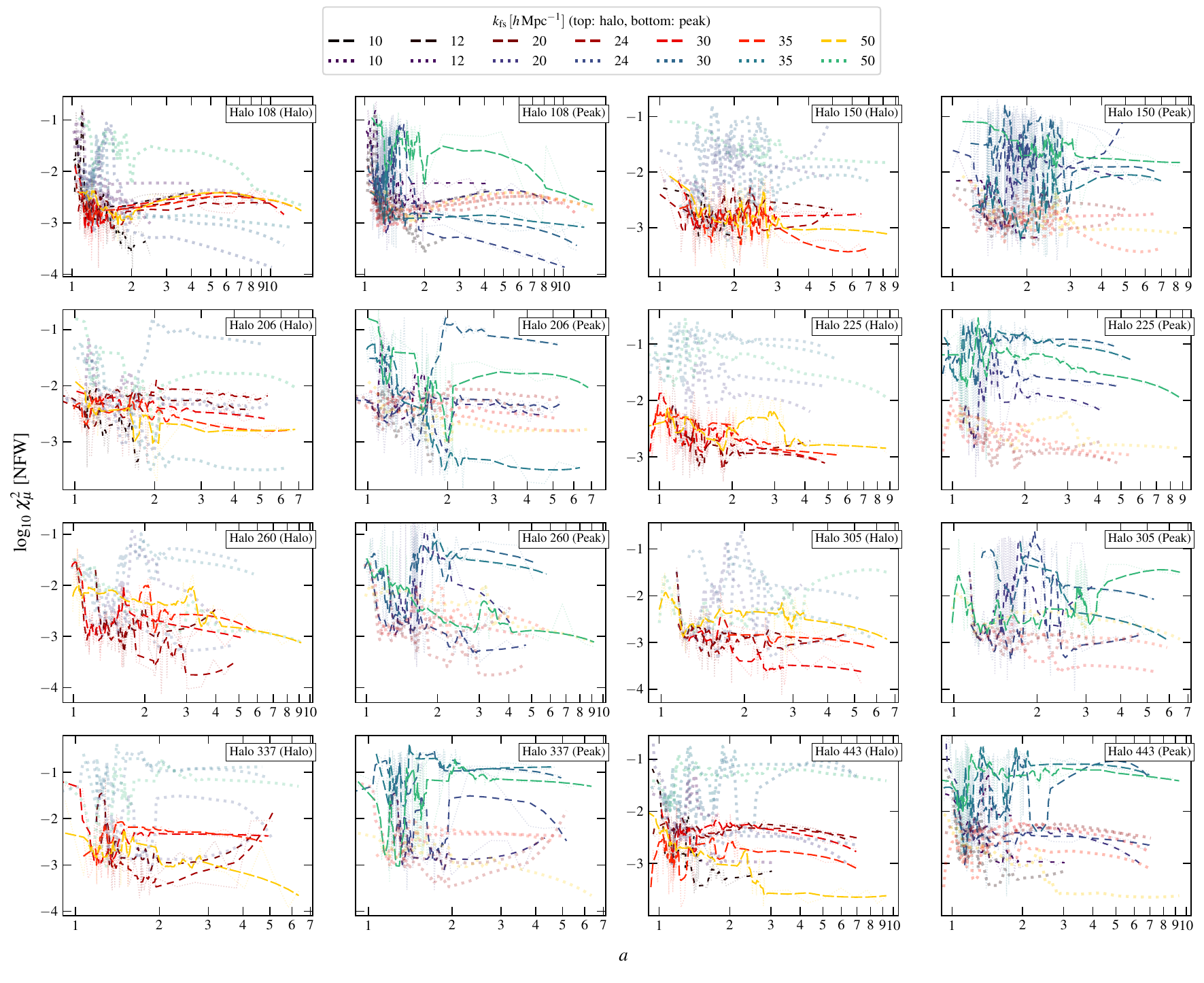}
    \caption{
    Evolution of the reduced chi-squared, $\chi_\mu^2$, of NFW fits to the inner density profiles as a function of the rescaled scale factor
    $a$. Panels are arranged in a $4\times 4$ grid, with each row showing a pair of haloes: the left and right panels in each row display the full halo density profile and the peak-only density profile respectively. In every panel, both components are shown for direct comparison: the components corresponding to that panel (halo or peak) are plotted in colour dashed line, while the complementary component are overplotted as a faint grey background curves, providing a visual reference for how the fit quality changes when the accreted material is removed.
    The halo identifier and the component type (``Halo'' or ``Peak'') are indicated in a boxed label at the upper right corner of each panel. For each component, curves are coloured according to the free-streaming wavenumber $k_{\rm fs}$, with distinct colormaps used for halo and peak profiles: halo curves span a \textit{red-to-yellow} colour gradient, while peak curves span a \textit{purple-to-green} gradient. Different line styles encode individual $k_{\rm fs}$ values within each colormap. For clarity, the $\chi_\mu^2$ measurements are smoothed with a Savitzky–Golay filter. This layout highlights both the time evolution of the NFW fit and the systematic difference between the full halo and peak-only density profiles.}
    \label{fig:NFWchi2}
\end{figure*}

Figure~\ref{fig:NFW_32} presents the evolution of the AICc difference, $\mathrm{AIC}\mathrm{c}[{\rm NFW}] - \mathrm{AIC}\mathrm{c}[{3/2}]$ for all haloes. The panels are arranged in the same $4\times 4$ layout as before, with halo and peak components shown side by side and the halo identifier and component type indicated in the upper-right corner. For each component, curves are coloured by $k_{\rm fs}$, using a red–to–yellow gradient for haloes and a purple–to–green gradient for peak-only profiles; different line styles distinguish individual $k_{\rm fs}$ values. A vertical light-grey bands mark snapshots where a reliable NFW fit could not be obtained and the inner profile is best regarded as a pure power law.

This suggests that prompt cusps are apparently destroyed very rapidly after collapse, consistent with our concerns about the impact of artificial fragments in the simulations; only haloes~206, 224 and 403 exhibit a short-lived phase during which a power-law profile is clearly favoured. 

In haloes~443, 305, 225, 150, and 337, the relative preference between NFW and the $-3/2$ model is more stable for the peak-only profiles than for the full haloes, indicating that once sufficient material has been added at large radii, the central remnant of the disturbed cusp is no longer further disrupted. There are a few cases in which the peak-only profile shows a similar preference for NFW as the full halo, such as haloes~260 and 150, which likely reflect major mergers that heavily disturb the central structure.

As $k_{\rm fs}$ increases and more substructure appears, one would expect the central cusp to be disrupted earlier. However, because artificial fragments can merge with the main halo immediately after collapse, the expected ordering with $k_{\rm fs}$ can be erased, producing an inverted behaviour such as that observed in haloes~260 and 305.

\begin{figure*}
    \hspace*{-1cm}
    \centering
    \includegraphics[width = 1.1\textwidth]{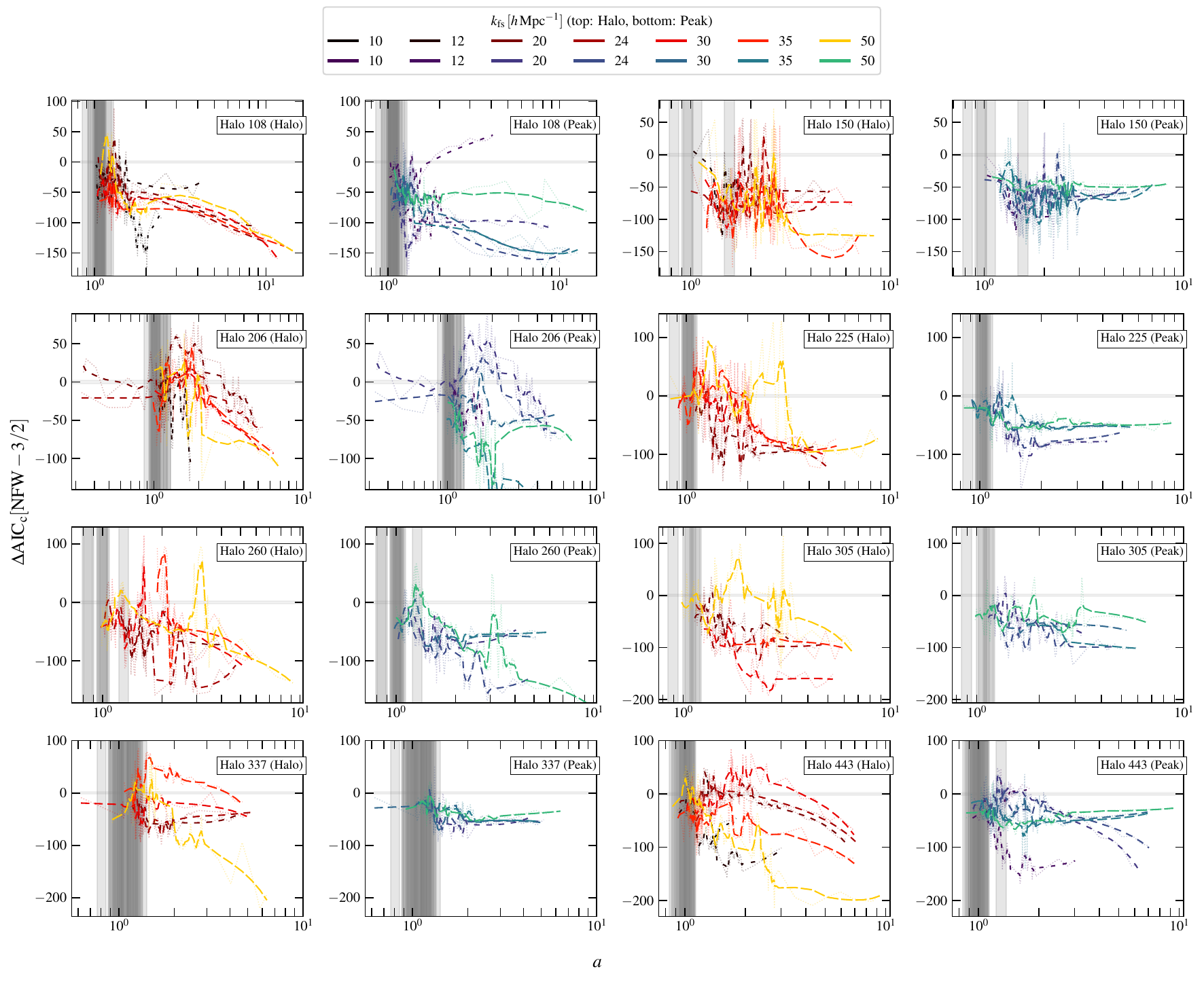}
    \caption{Model comparison between NFW and fixed-slope $\rho \propto r^{-3/2}$ profiles for the inner density structure of the haloes. Each panel shows the evolution of the AICc difference $\Delta\mathrm{AIC}_\mathrm{c} \equiv \mathrm{AIC}_\mathrm{c}[{\rm NFW}] - \mathrm{AIC}_\mathrm{c}[{3/2}]$ as a function of the rescaled scale factor $a$. The figure is arranged as a $4\times 4$ grid: in each row a pair of haloes
    is shown, with the left and right panels corresponding to the full halo density profile and the peak-only density profile respectively. A boxed label at the upper right of each panel indicates the halo identifier and whether the panel refers to the halo or peak component. Positive values $\Delta\mathrm{AIC}_\mathrm{c}$ indicate that the $-3/2$
    profile is statistically preferred over NFW, while negative values favour NFW. A horizontal grey band marks the region
    $|\Delta\mathrm{AIC}_\mathrm{c}|\leq 2$, within which the two models are statistically indistinguishable. Vertical light-red bands denote snapshots where a reliable NFW fit could not be obtained due to fitting failures; these intervals can be interpreted as prompt-cusp phases in which the inner structure is more cleanly described by a power-law density profile.
    }
    \label{fig:NFW_32}
\end{figure*}

In Figure~\ref{fig:32_Size} we briefly discuss the characteristic scale of the prompt cusps \(B\) denoting the number of inner logarithmic bins of the best fitting power law model. The inferred prompt cusp radii typically correspond to \(f_{\rm m}<0.5\). Rare cases approach (e.g. Halo 225, \(k_{\rm fs}=50\)), although we caution that this may reflect fitting noise given the persistently small \(r_B\) prior to a rapid rise around \(a\sim 2\). The behaviour of the peak only profiles closely parallels that of the halo profiles.
\begin{figure*}
    \hspace*{-1cm}
    \centering
    \includegraphics[width = 1.1\textwidth]{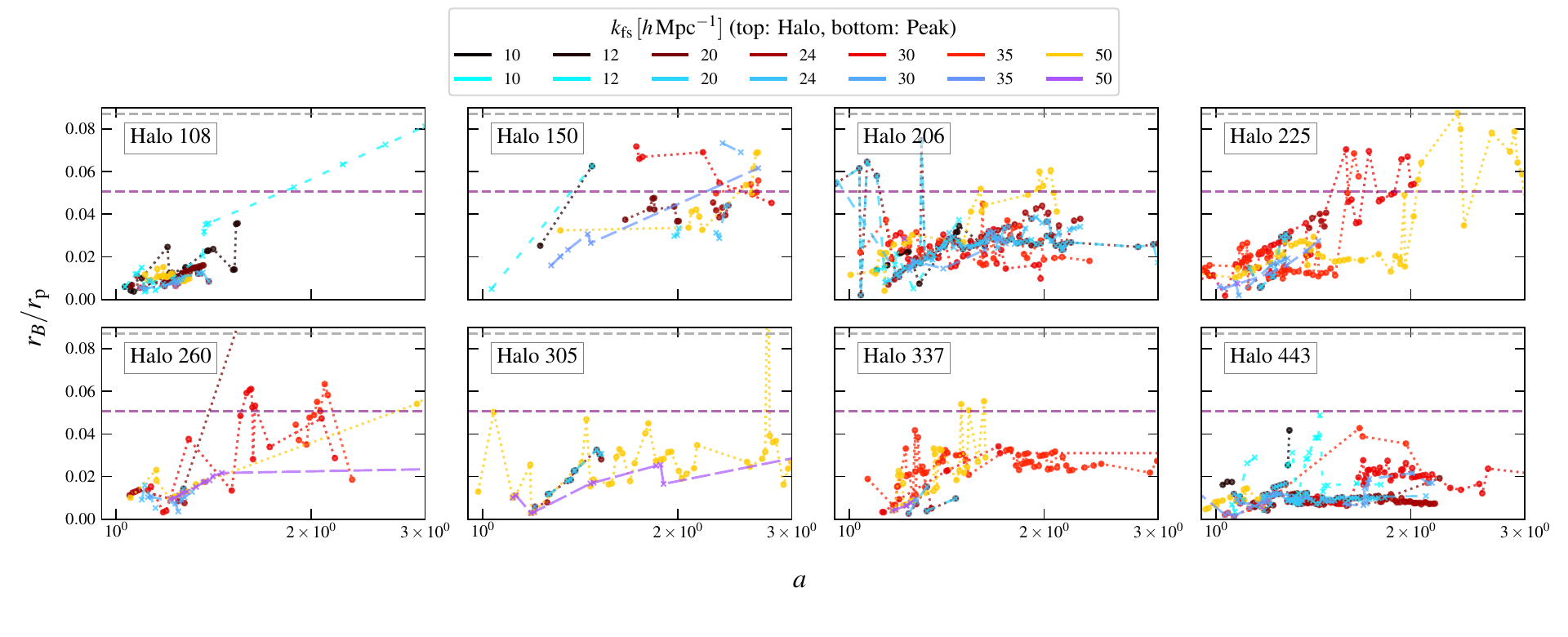}
    \caption{Evolution of the characteristic inner scale during the prompt-cusp phase. For each halo, we show $r_B/r_{\rm p}$ versus the rescaled time $a$, where \(r_B\) is the physical radius associated with the best-fitting inner-bin count \(B\) of the power-law \(r^{-3/2}\) fit (Sec.~\ref{subsec:fitting}) and \(r_{\rm p}\) is the size of the matched primordial peak (Sec.~\ref{subsec:self-similar}), restricted to epochs where $\Delta\mathrm{AIC}_\mathrm{c}[{\rm NFW} -{3/2}]>0$ and the epochs where the NFW fit fails (i.e. the $-3/2$ power law outperforms NFW). In each panel, the results corresponding to the halo profile is coloured from red to yellow with dotted lines and circle markers, while the peak profile results are plotted in blue-to-purple with dashed lines and $\times$ markers. Colours encode the different $k_{\rm sf}$ value, as indicated in the legend on the top. The text boxes at the upper left annotate the halo identifiers. Two horizontal dashed reference lines mark the theoretical radii where the enclosed mass equals the full peak mass (\textit{grey}) and half the peak mass (\textit{purple}).
    }
    \label{fig:32_Size}
\end{figure*}

Figures~\ref{fig:theory_compare_halo} shows, for halo profiles, $\Delta\mathrm{AIC}_\mathrm{c}[3/2{-}12/7]\equiv \mathrm{AIC}_\mathrm{c}[{3/2}]-\mathrm{AIC}_\mathrm{c}[{12/7}]$ as teal “$+$” symbols and $\Delta\mathrm{AIC}_\mathrm{c}[12/7{-}\lambda]\equiv \mathrm{AIC}_\mathrm{c}[{12/7}]-\mathrm{AIC}_\mathrm{c}[{\lambda}]$ as salmon “$\times$”, plotted against the rescaled time $a$. The information is similar to Fig.~\ref{fig:theory_compare_peak} discussed in Sec.~\ref{subsubsec:models}.

\begin{figure}
    \centering
    \includegraphics[width=\columnwidth]{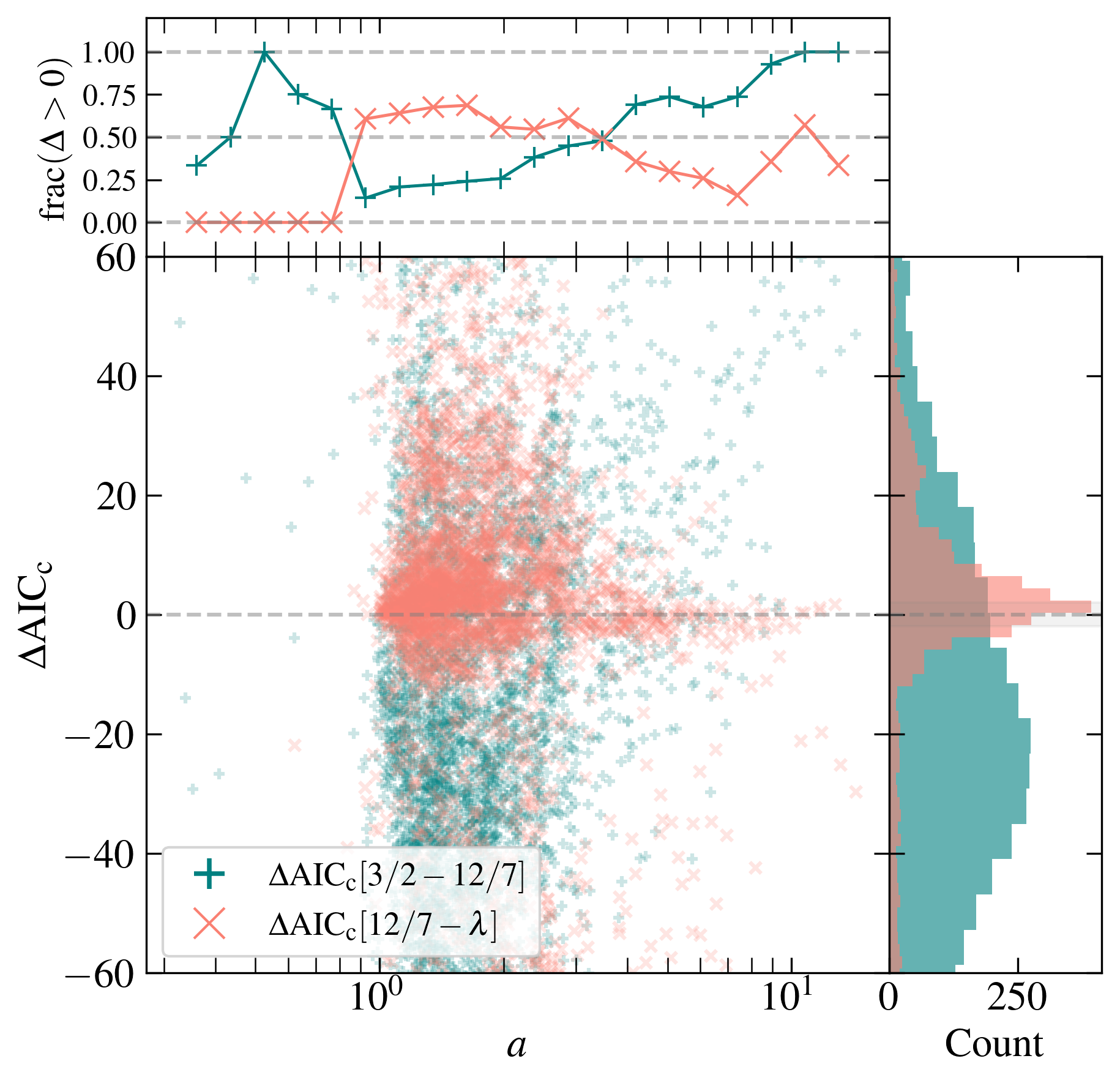}
    \caption{Model comparison of the \textit{halo} density profile across all haloes, snapshots, and $k_{\rm fs}$, restricted to epochs where $\Delta\mathrm{AIC}_\mathrm{c}[{\rm NFW} -{3/2}]>0$ (i.e. the $-3/2$ power law outperforms NFW), together with epochs where the NFW fit fails. Main panel shows $\Delta\mathrm{AIC}_\mathrm{c}[3/2-12/7]$ as $+$ and $\Delta\mathrm{AIC}_\mathrm{c}[12/7-\lambda]$ as $\times$ with \textit{x} axis as $a$ where $a$ is the rescaled scale factor. The top left panels shows the fraction of $\Delta\mathrm{AIC}_\mathrm{c}[M_1 - M_2] > 0$, with same colour and sign as the main panel. 
    The shaded band marks $|\Delta\mathrm{AIC}_\mathrm{c}|\leq 2$ where models are statistically indistinguishable. The side panel shows the distributions of the two $\Delta\mathrm{AIC}_\mathrm{c}$ samples (horizontal histograms) on the same $y$-axis scale as the main panel. 
    }
    \label{fig:theory_compare_halo}
\end{figure}


\bsp	
\label{lastpage}
\end{document}